\documentclass[10pt,journal,twoside]{IEEEtran}

\usepackage[utf8]{inputenc}   
\usepackage[T1]{fontenc}      

\usepackage{xcolor}

\usepackage{amsmath}
\usepackage{amsfonts}
\usepackage{mathtools}
\usepackage[cmintegrals]{newtxmath}
\usepackage{cancel}
\usepackage{balance}
\usepackage{stfloats}

\usepackage{graphicx}     
\usepackage{subfigure}   
\usepackage[update,prepend]{epstopdf} 

\usepackage{pgfplots}
\pgfplotsset{compat=1.18} 

\usepackage{tikz}
\usetikzlibrary{arrows.meta, positioning, shapes.geometric, decorations.pathmorphing, calc, angles,quotes}

\usepackage{cite}
\usepackage{setspace}

\usepackage{algorithm}
\usepackage{algpseudocode}

\usepackage{array}
\usepackage{booktabs}
\usepackage{threeparttable}
\usepackage{multirow}

\usepackage{caption}
\renewcommand{\arraystretch}{1.1}

\usepackage{siunitx}
\DeclareSIUnit[number-unit-product = {}]{\molecules}{\text{molecules}}

\usepackage{indentfirst} 
\usepackage{float}       

\usepackage{bbm,dsfont}

\newcommand{\argmax}{\operatornamewithlimits{argmax}}
\newcommand{\argmin}{\operatornamewithlimits{argmin}}

\newtheorem{remark}{Remark}

\usepackage[
    colorlinks=true,
    linkcolor=blue,
    citecolor=blue,
    urlcolor=blue
]{hyperref}

\begin{document}

\title{Field-Assisted Molecular Communication: Girsanov-Based Channel Modeling and Dynamic Waveform Optimization}

\author{Po-Chun~Chou, Yen-Chi~Lee,~\IEEEmembership{Member,~IEEE,} Chun-An~Yang,~\IEEEmembership{Graduate~Student~Member,~IEEE,} Chia-Han~Lee,~\IEEEmembership{Member,~IEEE,}~and~Ping-Cheng~Yeh,~\IEEEmembership{Member,~IEEE}

\thanks{This work was supported by the National Science and Technology Council of Taiwan (NSTC 113-2115-M-008-013-MY3). (Corresponding author: Yen-Chi Lee.)}
\thanks{P.-C. Chou and P.-C. Yeh are with the Graduate Institute of Communication Engineering, National Taiwan University, Taipei, Taiwan. (e-mail: \texttt{d02942012@ntu.edu.tw, pcyeh@ntu.edu.tw})}
\thanks{Y.-C. Lee is with the Department of Mathematics, National Central University, Taoyuan, Taiwan. (e-mail: \texttt{yclee@math.ncu.edu.tw})}%
\thanks{C.-A. Yang is with the Department of Computer Science, National Tsing Hua University, Hsinchu, Taiwan. (e-mail: \texttt{s110065505@m110.nthu.edu.tw})}%
\thanks{C.-H. Lee is with the Institute of Communications Engineering, National Yang Ming Chiao Tung University, Hsinchu, Taiwan. (e-mail: \texttt{chiahan@nycu.edu.tw})}
}

\maketitle

\begin{abstract}
Analytical modeling of field-assisted molecular communication under dynamic electric fields is fundamentally challenging due to the coupling between stochastic transport and complex boundary geometries, which renders conventional partial differential equation (PDE) approaches intractable. 
In this work, we introduce an effective stochastic modeling approach to address this challenge. By leveraging trajectory-reweighting techniques, we derive analytically tractable channel impulse response (CIR) expressions for both fully-absorbing and passive spherical receivers, where the latter serves as an exact theoretical baseline to validate our modeling accuracy. 
Building upon these models, we establish a dynamic waveform design framework for system optimization. Under a maximum \textit{a posteriori} decision-feedback equalizer (MAP-DFE) framework, we show that the first-slot received probability serves as the primary determinant of the bit error probability (BEP), while inter-symbol interference manifests as higher-order corrections.
Exploiting the monotonic response of the fully-absorbing architecture and using the limitations of the passive model to justify this strategic focus, we reformulate BEP minimization into a distance-based optimization problem. We propose a unified, low-complexity Maximize Received Probability (MRP) algorithm, encompassing the Maximize Hitting Probability (MHP) and Maximize Sensing Probability (MSP) methods, to dynamically enhance desired signals and suppress inter-symbol interference. Numerical results validate the accuracy of the proposed modeling approach and demonstrate near-optimal detection performance.
\end{abstract}

\begin{IEEEkeywords}
Field-assisted molecular communication, time-varying drift, active receiver, inter-symbol interference (ISI) mitigation, waveform optimization.
\end{IEEEkeywords}

\section{Introduction}
\label{sec:intro}

\IEEEPARstart{R}{ecent} advances in nanotechnology have enabled the development of nanomachines which, despite individual physical constraints, can achieve complex objectives through cooperative nanonetworks \cite{akyildiz_2008_nanonetwork}, necessitating effective paradigms such as molecular communication (MC) \cite{nakano_2013_Molecular_Communication, Farsad:2016}. Among various MC mechanisms—including diffusion \cite{pierobon2011noise, kadloor_2012_molecular}, gap junctions \cite{nakano2008molecular, bicen2016linear}, and molecular or bacterial motors \cite{moore2006design, gregori2010new}—molecular communication via diffusion (MCvD) is the most extensively investigated and serves as the focus of this work. To ensure reliable, \textit{time-slotted} information exchange, accurate end-to-end channel models encompassing emission, diffusion, and reception are essential \cite{pierobon2010physical}. In this study, we leverage existing techniques to assume perfect synchronization between the transmitter (Tx) and receiver (Rx) \cite{training_2017_hsu, syn_2018_Luo}, focusing our analysis on the underlying channel dynamics.

A critical consequence of these diffusion-driven dynamics is the inherent channel memory, where residual molecules from previous intervals cause inter-symbol interference (ISI) and degrade performance. While various strategies have been proposed to mitigate ISI—ranging from advanced detection filters \cite{Burcu_2015_ISI, pierobon2010physical} and chemical-based solutions \cite{Nasiri_2016_Type_based, 2014_Optimal_rx} to environmental flow-based approaches \cite{noel2014diffusive, Cyclostationary_Drift_2017_Orlik}—another line of research explores using external assistive fields to actively regulate transport. For instance, constant electric \cite{electric_2019} or magnetic fields \cite{wicke2019magnetic} can accelerate molecular propagation. However, fixed-field designs are limited in time-slotted systems because channel memory effects vary across intervals. This motivates the design of assistive fields as controllable, time-varying functions that can be dynamically shaped to match the underlying channel dynamics.

In biomedical communication scenarios, information-bearing particles are often ions, which naturally facilitates transport regulation via assistive electric fields \cite{byrne2015local}. In this work, we investigate such electrically assisted MCvD systems by incorporating two representative models \cite{jamali2019channel}: the passive (PA) and the fully-absorbing (FA) spherical receivers. The PA model, which yields an exact closed-form solution under our framework, is explicitly included as a rigorous benchmark to establish mathematical consistency and to theoretically justify why subsequent system optimizations must prioritize the FA architecture. While characterizing the channel impulse response (CIR) is essential for performance evaluation, the interplay between spherical boundary conditions and time-varying electric fields—described by the Nernst--Planck equation—renders conventional partial differential equation (PDE) approaches analytically intractable \cite{risken1989fokker, redner2001guide}. This shift from simpler cubic geometries \cite{Chou2022} to physically representative spherical models induces nontrivial boundary-value problems that create a significant analytical bottleneck. To overcome this, it is necessary to derive fundamentally new received probability expressions, which can then serve as the basis for robust symbol recovery under detection frameworks such as the maximum \textit{a posteriori} (MAP) rule paired with a decision-feedback equalizer (DFE) \cite{proakis2008digital, kislal2020isi}.

To overcome this analytical bottleneck and facilitate system-level design, we employ a stochastic framework that bypasses the direct solution of complex PDEs by reweighting particle trajectories \cite{lee2024characterizing, Lee2025Exact3D}. Rather than tackling the Nernst--Planck equation head-on, this approach leverages advanced stochastic techniques \cite{oksendal2013stochastic} to evaluate probability measures, effectively capturing the influence of time-varying electric fields as a reweighting of free-diffusion paths. By applying this trajectory-reweighting technique, we derive analytical tractable CIR expressions for both PA and FA spherical receivers.
These expressions are then subsequently integrated into a time-slotted MCvD system model, enabling a rigorous performance analysis under the MAP-DFE framework.

Beyond performance characterization, our analysis identifies the received probability as the primary determinant of system bit error probability (BEP). This insight allows us to reformulate the complex BEP minimization task into a more tractable objective: maximizing the received probability through the optimal design of the time-varying assistive electric field. Specifically, we propose two receiver-specific strategies—\textit{Maximize Sensing Probability (MSP)} for PA receivers and \textit{Maximize Hitting Probability (MHP)} for FA receivers—both of which are executed through a unified \textit{Maximize Received Probability (MRP)} algorithm. By mapping the optimization onto a distance-based optimization formulation, the MRP algorithm achieves near-optimal performance with remarkably low computational complexity, offering a scalable solution for various field-assisted molecular communication scenarios.

The main contributions of this work are summarized as follows.
\begin{itemize}
    \item[(i)] \textbf{Tractable Channel Modeling under Dynamic Drift:} We propose a novel stochastic framework leveraging trajectory-reweighting techniques to circumvent the analytical intractability of Nernst--Planck PDEs. By introducing the \textit{Effective Drift Approximation (EDA)}, we decouple the time-varying composite drift from complex spherical boundaries, deriving analytically tractable CIR expressions via effective drift mapping for fully-absorbing receivers, and exact CIR expressions for passive ones.
    \item[(ii)] \textbf{Fundamental Insights into Receiver Dynamics:} Building on this framework, we further show that a critical physical distinction exists between receiver architectures. We mathematically demonstrate that the fully-absorbing receiver exhibits a robust, monotonic response to assistive fields, ensuring a stable optimization landscape. Conversely, the passive receiver suffers from intrinsic physical limitations. By contrasting these behaviors against our exact baseline, we mathematically justify the strategic focus on active architectures for reliable nanonetwork design and subsequent transceiver optimization.
    \item[(iii)] \textbf{Unified Waveform Optimization (MRP Algorithm):} Recognizing that the received probability dominates the system BEP, we reformulate the intractable BEP minimization into a distance-based optimization problem aimed at maximizing the received probability. We propose the MRP algorithm, which intelligently allocates the energy budget into a two-phase waveform: a signal-enhancing acceleration phase and an ISI-suppressing counter-drift phase.
    \item[(iv)] \textbf{Near-Optimal Performance with Low Complexity:} Extensive numerical evaluations validate the accuracy of our derived CIRs. Under a MAP-DFE detection framework, the lightweight MRP algorithm achieves near-optimal BEP performance, significantly outperforming conventional constant-field baselines.
\end{itemize}

The remainder of this paper is organized as follows. Section~\ref{sec:system_model} establishes the physical system model and introduces the measure-theoretic stochastic framework for characterizing particle dynamics under time-varying drift. Section~\ref{sec:analytical_CIR} derives the closed-form analytical CIR expressions for both fully-absorbing and passive spherical receivers. Section~\ref{sec:DTsignal_model} presents the discrete-time signal model and investigates the fundamental physical properties of the channel coefficients, specifically highlighting the distinction between receiver architectures. Section~\ref{sec:BEP_analysis} details the symbol detection rule and the evaluation of BEP under the MAP-DFE framework. Section~\ref{sec:wf_optimization} formulates the waveform optimization problem and introduces the unified MRP algorithm for assisted electric field design. Numerical results and validations are presented in Section~\ref{sec:numerical}, and finally, Section~\ref{sec:conclude} concludes the paper and summarizes the key findings.

\section{System Model and Stochastic Formulation}\label{sec:system_model}

To characterize the end-to-end behavior of field-assisted molecular communication (FAMC) systems, it is necessary to establish a tractable stochastic model that encapsulates both the physical receiver geometry and the particle dynamics under a time-varying drift. In this section, we define and set up the physical environment and introduce a measure-theoretic framework that serves as the universal analytical engine for our channel derivations.

\begin{figure}[!t]
    \centering
    \definecolor{myBlue}{RGB}{0, 50, 150}
    \definecolor{myPurple}{RGB}{120, 0, 120}
    \definecolor{myGreen}{RGB}{0, 120, 0}
    \definecolor{myRed}{RGB}{200, 0, 0}

    \resizebox{\columnwidth}{!}{
        \begin{tikzpicture}[
            >=Stealth, 
            font=\small,
            rx_color/.style={color=myBlue},
            tx_color/.style={color=myPurple},
            field_color/.style={color=myGreen},
            dim_color/.style={color=myRed}
        ]

            \pgfmathsetseed{3}
        
            \coordinate (O) at (0,0);
            \draw[->, thick, myBlue] (O) -- (2.5,0) node[right] {$x_1$-axis};
            \draw[->, thick, myBlue] (O) -- (0,2) node[above] {$x_2$-axis};
            \draw[->, thick, myBlue] (O) -- (-1.2,-1.2) node[below left] {$x_3$-axis};
            \node[myBlue, anchor=south west, xshift=2pt] at (O) {\footnotesize Origin};
        
            \draw[thick, dashed, myBlue] (O) circle (1.2cm);
            \node[myBlue, font=\bfseries] at (1.8, 1.3) {Spherical Rx};
            \draw[<->, myBlue!70!black, thick] (O) -- (-0.85, 0.85) node[midway, left, xshift=-2pt] {$r_{\text{Rx}}$};
        
            \coordinate (Tx) at (-5, 0);
            \fill[myPurple] (Tx) circle (2.5pt);
            \node[black, font=\small, anchor=south] at (-5, 0.2) {Point Source Tx at $\mathbf{x}_0$};
        
            \coordinate (TargetP) at (-4.2, -0.6);
            
            \draw[->, myPurple, dashed, thick, decoration={snake, amplitude=1.2pt, segment length=6pt}, decorate] 
                (Tx) -- (-4.8, -0.4) -- (TargetP);
        
            \fill[myPurple] (TargetP) circle (1.8pt);
        
            \foreach \i in {1,...,7} {
                \fill[myPurple!70] ($(Tx) + (rand*0.8 + 0.5, rand*0.6 - 0.2)$) circle (1.8pt);
            }
            
            \node[myPurple, font=\bfseries] at (-5, -1.5) {Message Particles};
        
            \draw[->, myGreen, dashed, line width=1.5pt] (-3.5, 3.2) -- (-1.5, 3.2) 
                node[midway, above, font=\bfseries] {Composite Drift $\boldsymbol{\Phi}(t)$};
        
            \draw[<->, myRed, thick] (-5, 2.5) -- (0, 2.5) node[midway, above] {$d_0 = \|\mathbf{x}_0\|$};
            \draw[myRed, thin, dashed] (-5, 0) -- (-5, 2.6);
            \draw[myRed, thin, dashed] (0, 0) -- (0, 2.6);
        
        \end{tikzpicture}
    }
    \caption{System illustration of MCvD with a time-varying assisted electric field. A point Tx located at $\mathbf{x}_0$ releases charged particles, while a spherical Rx centered at the origin captures the signal either passively or via full absorption. The trajectories are influenced by a composite drift field $\boldsymbol{\Phi}(t)$, combining periodic background flow with an externally designed electric field aligned along the $x_1$-axis.}
    \label{fig:system_model}
\end{figure}
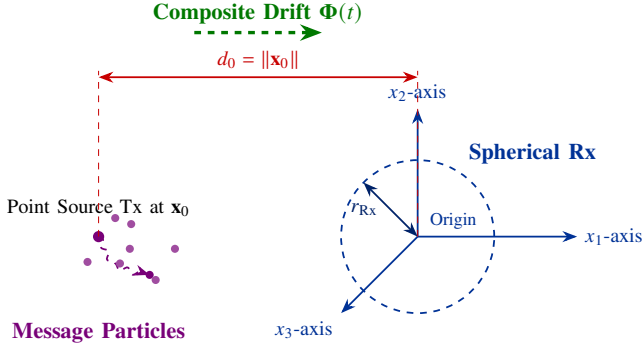

\subsection{Physical System and Receiver Models}

We consider a molecular communication system operating in a three-dimensional (3-D) fluidic environment, as illustrated in Fig.~\ref{fig:system_model}. The system comprises a point Tx located at $\mathbf{x}_0$ and a spherical Rx of radius $r_{\mathrm{Rx}}$ centered at the origin. The Euclidean distance from the Tx to the Rx center is denoted by $d_0 = \|\mathbf{x}_0\|$.

Communication is analyzed in a continuous-time framework starting from an initial emission. At $t=0$, the Tx impulsively releases a designated quantity of $N$ charged message particles (ions) into the fluidic environment. 
We focus our analysis on the underlying transport dynamics and the resulting reception statistics, which form the fundamental basis for subsequent discrete-time system modeling. For clarity, the key notations used throughout this paper are summarized in TABLE~\ref{tab:notations}.

To comprehensively characterize the reception process, we investigate two distinct spherical Rx models, ordered by their biological relevance and analytical complexity:
\begin{itemize}
    \item \textbf{Fully-Absorbing (Active) Rx:} Particles are permanently removed from the fluidic channel upon their first contact with the receiver surface, defined by 
    \begin{equation}
    \mathcal{R}_{\mathrm{hit}} = \{\mathbf{x} \mid \|\mathbf{x}\| = r_{\mathrm{Rx}}\}.
    \end{equation}
    This absorbing boundary accurately models the ligand-receptor binding mechanisms prevalent in targeted drug delivery and biological cells \cite{schulten2000lectures}. Due to its practical significance in bio-nanomachines, this model constitutes the primary focus of our system design.
    \item \textbf{Passive Rx:} Message particles can freely enter and exit the sensing volume 
    \begin{equation}
    \mathcal{R}_{\mathrm{s}} = \{\mathbf{x} \mid \|\mathbf{x}\| \leq r_{\mathrm{Rx}}\}
    \end{equation}
    without undergoing chemical reactions or physical absorption \cite{nelson2008biological}. While less representative of biological receptors, this ``transparent'' boundary serves as a fundamental theoretical baseline and a mathematical consistency check for our unified framework.
\end{itemize}

The propagation of the charged particles is governed by a 3-D composite drift field. This field consists of a periodic background flow $\mathbf{u}(t)$ \cite{Cyclostationary_Drift_2017_Orlik} and an externally controlled, time-varying assisted electric field $\mathbf{E}(t)$ \cite{electric_2019}. The composite drift field is defined as
\begin{align}
    \boldsymbol{\Phi}(t) := \mathbf{u}(t) + c_e \mathbf{E}(t), \label{eq:composite_drift}
\end{align}
where $c_e = \mu z e$ is the electrophoretic mobility constant, with $\mu = \frac{D}{k_B T}$ representing the mechanical mobility, $D$ being the diffusion coefficient, $z$ the ion valence, $e$ the elementary charge, $k_B$ the Boltzmann constant, and $T$ the absolute fluid temperature. In this generalized framework, the spatially uniform composite drift $\boldsymbol{\Phi}(t)$ is treated as an arbitrary time-varying vector field, enabling the characterization of particle dynamics under complex, non-stationary environmental flows and assistive electric fields.

\begin{table}[!t] 
\caption{Key Notations}
\label{tab:notations}
\centering
\renewcommand{\arraystretch}{1.1} 
\begin{tabular}{p{0.32\columnwidth} p{0.58\columnwidth}}
\toprule
\textbf{Symbol} & \textbf{Description} \\
\midrule
$r_{\mathrm{Rx}}, \mathcal{R}_{\mathrm{s}}, \mathcal{R}_{\mathrm{hit}}$ & Rx radius, sensing region, and absorbing boundary. \\
$D, c_e$ & Diffusion coefficient and mobility constant. \\
$\boldsymbol{\Phi}(t), \mathbf{E}(t)$ & Composite drift and assisted electric field. \\
$\xi, \beta$ & Total energy budget and suppression onset fraction. \\
$\mathbb{P}, \mathbb{Q}, M_t$ & Reference/physical measures and likelihood ratio. \\
$\boldsymbol{\Phi}_{\mathrm{eff}}(T)$ & Time-averaged effective drift field over $[0, T]$. \\
$f_{\mathrm{hit}}^{\boldsymbol{\Phi}}(t), p_{\mathrm{s}}(T)$ & FHT density (active) and sensing prob. (passive). \\
$T_b, t_s, t_{\mathrm{peak}}$ & Symbol duration, sampling time, and peak time. \\
$p_{\mathrm{R}}[i]$ & Received probability in the $i$-th time slot. \\
$V_1, V_2, q_{\mathrm{S}}, q_{\mathrm{R}}$ & Field magnitudes and ideal velocities for signal/ISI.\\
\bottomrule
\end{tabular}
\end{table}

\subsection{Stochastic Particle Dynamics under Time-Varying Drift}

The concentration evolution of the charged particles under the Nernst--Planck equation can be translated into the stochastic trajectory of a single ion via a stochastic differential equation (SDE). The It\^o process governing the position $\mathbf{X}_t$ of a single ion at time $t$ is given by
\begin{align}
    d\mathbf{X}_t = \boldsymbol{\Phi}(t) dt + \sqrt{2D} \, d\mathbf{B}_t, \label{eq:ito_diff}
\end{align}
where $\mathbf{B}_t$ is a standard 3-D Brownian motion defined on a probability space $(\Omega, \mathcal{F}, \mathbb{P})$ with $\mathbf{B}_0 = \mathbf{0}$, and $\boldsymbol{\Phi}(t)$ acts as the deterministic, time-varying drift.

Evaluating the first-hitting time (FHT) density for the absorbing boundary under a time-varying drift directly from the Fokker--Planck PDE is notoriously intractable. To bypass this barrier and facilitate subsequent system-level optimization, we employ an effective stochastic modeling approach leveraging advanced trajectory-reweighting techniques \cite{oksendal2013stochastic,bhattacharya2009stochastic}. This framework allows us to analytically decouple any arbitrary time-varying drift vector from the complex boundary geometry to obtain tractable channel coefficients.

Specifically, rather than solving the PDE directly, we consider the process under a reference measure $\mathbb{P}$ where it behaves as a pure diffusion (zero-drift) process, i.e., $d\mathbf{X}_t = \sqrt{2D}\, d\mathbf{B}_t$. Since the physical drift $\boldsymbol{\Phi}(t)$ is deterministic and bounded, it satisfies the Novikov condition \cite{oksendal2013stochastic}. According to the CMG theorem, there exists an equivalent probability measure $\mathbb{Q}$ under which the process exhibits the true time-varying drift. The transformation between the physical measure $\mathbb{Q}$ and the reference measure $\mathbb{P}$ is governed by the Radon--Nikodym derivative (i.e., the likelihood ratio process),
\begin{align}
    M_t = \frac{d\mathbb{Q}}{d\mathbb{P}} = \exp\left( \int_0^t \frac{\boldsymbol{\Phi}(\alpha)}{\sqrt{2D}} \cdot d\mathbf{B}_\alpha - \frac{1}{2} \int_0^t \frac{\|\boldsymbol{\Phi}(\alpha)\|^2}{2D}\, d\alpha \right). \label{eq:Radon_Nikodym_density_case}
\end{align}

Let $A$ denote the arbitrary event of a particle arriving at a target spatial state $\mathbf{x}$ at observation time $T$. Utilizing the EDA (see Appendix~\ref{appendix:EDA_approx}) over the interval $[0, T]$, where the effective drift is defined as the time-average $\boldsymbol{\Phi}_{\mathrm{eff}}(T) := \frac{1}{T} \int_0^T \boldsymbol{\Phi}(\alpha) d\alpha$, we can evaluate the probability of event $A$ under the physical measure $\mathbb{Q}$ by evaluating the stochastic integral using the endpoint $\mathbf{X}_T - \mathbf{x}_0 = \sqrt{2D}\mathbf{B}_T$,
\begin{align}
    \mathbb{Q}(A) &= \mathbb{E}^{\mathbb{P}} \left[ M_T \cdot \mathbbm{1}_A \right] \nonumber \\
    &\approx \exp\left( \frac{\boldsymbol{\Phi}_{\mathrm{eff}}(T) \cdot (\mathbf{x} - \mathbf{x}_0)}{2D} - \frac{\|\boldsymbol{\Phi}_{\mathrm{eff}}(T)\|^2 T}{4D} \right) \cdot \mathbb{P}(A), \label{eq:radon_nikodym_event_approx}
\end{align}
where $\mathbb{P}(A)$ is the probability of the event under pure diffusion (zero drift), and the exponential term serves as a geometric weighting factor encapsulating the energy and directionality of the assisted electric field. The EDA should be interpreted as a path-averaged drift approximation. In particular, the approximation is most accurate when the variation of the drift is slow relative to the characteristic hitting time scale. This is empirically validated in Fig.~\ref{fig:verification}. Importantly, the approximation preserves the exponential tilting structure induced by the Girsanov transformation.

Expressed in terms of probability density functions, the joint spatial-temporal distribution $f^{\boldsymbol{\Phi}}(\mathbf{x}, T)$ under the time-varying electric field can be seamlessly derived from the zero-drift distribution $f(\mathbf{x}, T)$ via
\begin{equation}
    f^{\boldsymbol{\Phi}}(\mathbf{x}, T) = \exp\left( \frac{\boldsymbol{\Phi}_{\mathrm{eff}}(T) \cdot (\mathbf{x} - \mathbf{x}_0)}{2D} - \frac{\|\boldsymbol{\Phi}_{\mathrm{eff}}(T)\|^2 T}{4D} \right) f(\mathbf{x}, T). \label{eq:f_trans_pdf_final}
\end{equation}

This exponential tilting relationship \eqref{eq:f_trans_pdf_final} serves as our core analytical framework. In the subsequent section, we demonstrate how this framework elegantly resolves the CIR for both the formidable fully-absorbing boundary and the baseline passive sensing volume, providing a general mapping regardless of the relative orientation between the drift vector and the Tx--Rx axis.

\section{Analytical Channel Impulse Responses}\label{sec:analytical_CIR}

Building on the stochastic modeling framework developed in Section~\ref{sec:system_model}, this section evaluates the CIR for electrically assisted MCvD systems. The CIR encapsulates the probability that a single message particle, impulsively released at the transmitter at time $t=0$, is successfully captured or sensed at the receiver at observation time $T$. 

We first tackle the primary challenge of this work: characterizing the hitting density for a fully-absorbing boundary under a \emph{time-varying} electric field. Subsequently, we apply the identical measure-theoretic engine to the passive receiver to verify the theoretical consistency of our unified framework.

\subsection{CIR of the Fully-Absorbing Receiver}

For the FA (active) receiver, the CIR is defined by the FHT density. Unlike passive observation, absorption is a path-dependent process; if a particle touches the boundary $\mathcal{R}_{\mathrm{hit}}$ at any time $t \le T$, it is permanently removed from the fluidic channel. 

Recently, the exact analytical CIR for a FA spherical receiver under a \emph{constant} uniform drift $\mathbf{v}$ with an arbitrary direction was established in \cite{Lee2025Exact3D}. Let $f_{\mathrm{hit}}^{(\mathbf{v})}(t)$ denote this zero-varying-drift CIR. By utilizing the joint time-location distribution and a spatial measure change, it was shown in \cite{Lee2025Exact3D} that
\begin{IEEEeqnarray}{rCl}
    f_{\mathrm{hit}}^{(\mathbf{v})}(t) &=& \exp\biggl( -\frac{\mathbf{v}\cdot\mathbf{x}_0}{2D} -\frac{\|\mathbf{v}\|^{2}t}{4D} \biggr) \frac{\sqrt{4D}}{\sqrt{\pi}\|\mathbf{v}\|^{1/2}\|\mathbf{x}_0\|^{1/2}} \nonumber \\
    && \times \sum_{m=0}^{\infty} \left(m+\tfrac{1}{2}\right) I_{m+\frac{1}{2}}\!\left(\frac{\|\mathbf{v}\| r_{\mathrm{Rx}}}{2D}\right) P_m(\cos\psi) \nonumber \\
    && \times \int_{0}^{\infty} \mathcal{H}_m(\lambda)\, e^{-\frac{1}{2}\lambda^2 t}\,d\lambda,
    \label{eq:hit_constant_drift}
\end{IEEEeqnarray}
where $\psi = \angle(\mathbf{v}, \mathbf{x}_0)$ is the drift angle, $I_{\nu}(\cdot)$ is the modified Bessel function of the first kind, $P_m(\cdot)$ is the Legendre polynomial, and $\mathcal{H}_m(\lambda)$ is the geometry-dependent spectral kernel composed of cross-products of Bessel functions (detailed definitions can be found in \cite{Lee2025Exact3D}).

However, directly solving the Fokker--Planck equation or applying traditional boundary value methods for a \emph{time-varying} composite drift field $\boldsymbol{\Phi}(t)$ is analytically intractable. To bridge this critical gap, we leverage the effective drift mapping formulated in Section~\ref{sec:system_model}.

Under the EDA, the highly complex temporal variations of the electric field over the interval $[0, T]$ are integrated into the likelihood ratio process \eqref{eq:radon_nikodym_event_approx}. This essentially linearizes the path-dependent weighting, allowing us to decouple the time-varying field from the spherical absorbing boundary. Mathematically, this maps the time-varying problem into an equivalent constant-drift scenario, where the equivalent constant drift is precisely the time-averaged effective drift $\boldsymbol{\Phi}_{\mathrm{eff}}(T) = \frac{1}{T}\int_0^T \boldsymbol{\Phi}(\alpha) d\alpha$.

Consequently, the analytically tractable CIR via effective drift mapping under the time-varying electric field, denoted by $f_{\mathrm{hit}}^{\boldsymbol{\Phi}}(T)$, can be elegantly obtained by substituting the constant drift $\mathbf{v}$ in \eqref{eq:hit_constant_drift} with our effective drift vector $\boldsymbol{\Phi}_{\mathrm{eff}}(T)$. This substitution inherently constitutes a quasi-static approximation based on the EDA. By treating the effective drift as a constant representing the time-averaged path weighting over the observation interval, this approach maintains high computational efficiency while preserving analytical accuracy, particularly when the time-varying field changes relatively slowly compared to the characteristic hitting time. 
Namely, 
\begin{IEEEeqnarray}{rCl}
    f_{\mathrm{hit}}^{\boldsymbol{\Phi}}(T) &\approx& f_{\mathrm{hit}}^{(\mathbf{v})}\!\left(T\right) \Biggr|_{\mathbf{v} = \boldsymbol{\Phi}_{\mathrm{eff}}(T)} \nonumber \\
    &=& \exp\biggl( -\frac{\boldsymbol{\Phi}_{\mathrm{eff}}(T)\cdot\mathbf{x}_0}{2D} -\frac{\|\boldsymbol{\Phi}_{\mathrm{eff}}(T)\|^{2}T}{4D} \biggr) \nonumber \\
    && \times \frac{\sqrt{4D}}{\sqrt{\pi}\|\boldsymbol{\Phi}_{\mathrm{eff}}(T)\|^{1/2}\|\mathbf{x}_0\|^{1/2}} \nonumber \\
    && \times \sum_{m=0}^{\infty} \left(m+\tfrac{1}{2}\right) I_{m+\frac{1}{2}}\!\left(\frac{\|\boldsymbol{\Phi}_{\mathrm{eff}}(T)\| r_{\mathrm{Rx}}}{2D}\right) \nonumber \\
    && \times P_m(\cos\psi^{(\boldsymbol{\Phi})}) \int_{0}^{\infty} \mathcal{H}_m(\lambda)\, e^{-\frac{1}{2}\lambda^2 T}\,d\lambda,
    \label{eq:hit_time_varying_final}
\end{IEEEeqnarray}
where the effective drift angle is $\psi^{(\boldsymbol{\Phi})} = \angle(\boldsymbol{\Phi}_{\mathrm{eff}}(T), \mathbf{x}_0)$.

It is worth emphasizing that \eqref{eq:hit_time_varying_final} is a powerful and general 3-D analytical result. The effective drift $\boldsymbol{\Phi}_{\mathrm{eff}}(T)$ can point in any direction relative to the Tx--Rx axis, with its physical impact seamlessly captured through the time-averaged magnitude and the dynamic effective drift angle $\psi^{(\boldsymbol{\Phi})}$. This formulation proves that by using the Girsanov measure change, the impact of arbitrary time-varying assistive electric fields can be incorporated into the CIR as an explicit reweighting of the geometric modes, preserving analytical tractability for the subsequent system evaluation.

\subsection{CIR of the Passive Receiver}

To demonstrate the universality of our measure-theoretic approach, we apply the identical transformation engine to the passive receiver case. Unlike the active receiver, passive sensing evaluates the snapshot probability of a particle residing within the transparent sensing volume $\mathcal{R}_{\mathrm{s}}$ at observation time $T$. 
For mathematical convenience, we interchange the coordinate system such that the Tx is at the origin and the Rx is centered at $\mathbf{x}_0$. Under zero drift, the concentration profile is the fundamental solution of Fick's law \cite{berg1993random},
\begin{align}
    c(\mathbf{x};T) = \frac{1}{(4\pi DT)^{\frac{3}{2}}} \exp\!\left( -\frac{\|\mathbf{x}\|^{2}}{4DT} \right). \label{eq:sol_fick_law}
\end{align}

Applying the transformation relation \eqref{eq:f_trans_pdf_final}, the concentration profile under the time-varying drift $\boldsymbol{\Phi}(t)$ is reweighted by the Girsanov density,
\begin{align}
    c^{\boldsymbol{\Phi}}(\mathbf{x};T) = \exp\!\left( \frac{\boldsymbol{\Phi}_{\mathrm{eff}}(T)\cdot\mathbf{x}}{2D} -\frac{\|\boldsymbol{\Phi}_{\mathrm{eff}}(T)\|^{2}T}{4D} \right) c(\mathbf{x};T). \label{eq:concentration_tilt}
\end{align}
By completing the square in the exponent and recognizing that $T \boldsymbol{\Phi}_{\mathrm{eff}}(T) = \int_0^T \boldsymbol{\Phi}(\alpha) d\alpha$, the density seamlessly condenses into a shifted Gaussian form,
\begin{align}
    c^{\boldsymbol{\Phi}}(\mathbf{x};T) = \frac{1}{(4\pi DT)^{\frac{3}{2}}} \exp\!\left( -\frac{\|\mathbf{x}-T\boldsymbol{\Phi}_{\mathrm{eff}}(T)\|^{2}}{4DT} \right), \label{eq:concentration_shifted}
\end{align}
where $T\boldsymbol{\Phi}_{\mathrm{eff}}(T)$ denotes the cumulative drift vector representing the total spatial displacement.

\vspace{1ex}
\begin{remark}[Modeling Consistency]
It is crucial to note that for the passive observation case, our proposed effective drift mapping inherently degenerates to an \emph{exact} characterization.
\label{rmk:1}
\end{remark}
\vspace{1ex}

The CIR (i.e., sensing probability) of the PA spherical Rx is finally obtained by integrating \eqref{eq:concentration_shifted} over the transparent volume $\mathcal{R}_{\mathrm{s}}$ (the detailed step-by-step volumetric integration is provided in Appendix~\ref{sec:DSPforPass}),
\begin{align}
    p_{\mathrm{s}}(T) &:= \int_{\mathcal{R}_{\mathrm{s}}} c^{\boldsymbol{\Phi}}(\mathbf{x};T)\,d\mathbf{x} \nonumber\\
    &= \frac{1}{2} \!\left[ \mathrm{erf}\!\left( \frac{r_{\mathrm{Rx}}-\overline{d}_0}{\sqrt{4DT}} \right) + \mathrm{erf}\!\left( \frac{r_{\mathrm{Rx}}+\overline{d}_0}{\sqrt{4DT}} \right) \right] \nonumber\\
    &\quad - \frac{\sqrt{DT}}{\overline{d}_0\sqrt{\pi}} \!\left[ \exp\!\left( -\frac{(r_{\mathrm{Rx}}-\overline{d}_0)^2}{4DT} \right) - \exp\!\left( -\frac{(r_{\mathrm{Rx}}+\overline{d}_0)^2}{4DT} \right) \right], \label{eq:spherical_rx}
\end{align}
where $\overline{d}_0 := \bigl\|\mathbf{x}_0-T\boldsymbol{\Phi}_{\mathrm{eff}}(T)\bigr\|$ represents the effective Tx--Rx separation. This distance implicitly accounts for the total 3-D spatial displacement induced by the time-varying field, providing a robust, geometry-independent validation of our analytical framework.

\section{Discrete-Time Signaling and System Properties}\label{sec:DTsignal_model}

With the analytical CIRs established, we now transition from the continuous-time physical diffusion model to a discrete-time communication system. In this section, we formulate the end-to-end signal model and analyze the physical properties of the channel coefficients. Crucially, we identify a fundamental trade-off between receiver geometries that motivates our subsequent optimization strategy.

\subsection{Time-Slotted Transmission Model}

To enable successive symbol transmissions, we adopt a time-slotted MCvD framework \cite{2019_time_slotted_cho} with a symbol duration $T_b$. We consider \textit{On-Off Keying (OOK)} modulation (see \cite{shi2018error}), where the transmitted bit $b_i \in \{0, 1\}$ in slot $i$ dictates the emission process: if $b_i = 1$, an impulsive release of $N$ particles occurs at the start of the slot; if $b_i = 0$, no particles are released. 

For efficient transport and tractable waveform design, we henceforth assume that the assistive electric field $\mathbf{E}(t)$ is aligned with the Tx--Rx axis. Under this configuration, the composite drift vector can be characterized by its scalar component $\Phi(t)$ along the transmission axis. The received probability $p_\mathrm{R}[i]$ in the $i$-th time slot, originating from a single emission at $t=0$, is given by
\begin{IEEEeqnarray}{rCl}
    p_\mathrm{R}[i] = 
    \begin{cases}
        \int_{(i-1)T_b}^{iT_b} f_{\mathrm{hit}}^{\Phi}(t)\, dt, & \text{Fully-Absorbing Rx}; \\
        p_{\mathrm{s}}\bigl(t_s+(i-1)T_b\bigr), & \text{Passive Rx},
    \end{cases}
    \label{eq:received_prob_cases}
\end{IEEEeqnarray}
where $t_s \in (0, T_b]$ is the sampling time for the PA receiver. 

Assuming a sufficiently large $N$ (corresponding to dilute-solution regime), the total received ions $y[i]$ in slot $i$ follows a Gaussian distribution due to the linear superposition of the current signal and the ISI from previous slots \cite{Arrival_modelling_2014}, 
\begin{IEEEeqnarray}{rCl}
    y[i] = N b_i p_\mathrm{R}[1] + \sum^{i-1}_{k=1} N b_{i-k} p_\mathrm{R}[k+1] + N_{\text{total}}[i],
    \label{eq:received_signal_yi}
\end{IEEEeqnarray}
where $N_{\text{total}}[i]$ encapsulates the signal-dependent counting noise from all relevant slots. Equation \eqref{eq:received_signal_yi} highlights that the system performance is dictated by the target signal strength $p_\mathrm{R}[1]$ relative to the interference weights $p_\mathrm{R}[k+1]$ for $k \ge 1$.

\subsection{Physical Properties of the Channel Coefficients}

To minimize the BEP, the time-varying field $E(t)$ must be designed to maximize $p_\mathrm{R}[1]$ while suppressing ISI. However, the feasibility of this optimization strictly depends on the receiver's physical nature.

\vspace{1ex}
\begin{remark}[On the Robustness and Monotonicity of the Target Function]
While our unified Girsanov framework analytically captures both architectures, their responses to a directed drift field are fundamentally different.
\begin{itemize}
    \item \textbf{Fully-Absorbing Rx (Monotonicity):} As an active capture mechanism, the hitting probability is a cumulative process. This follows from the cumulative nature of absorption, where any acceleration of particle trajectories strictly increases the total hitting probability over time. Increasing the drift field strength strictly accelerates particle arrivals, causing $p_\mathrm{R}[1]$ to monotonically increase towards saturation. This property ensures a well-behaved, concave optimization landscape for gradient-based designs.
    \item \textbf{Passive Rx (Non-Monotonicity):} Sensing is a snapshot measurement within a transparent volume. Excessively strong drift fields can force the molecular cloud to traverse and exit the sensing volume before the sampling time $t_s$. This intrinsic physical limitation results in a non-monotonic ``bell-shaped'' response, introducing non-concave regions and severe numerical sensitivity into the optimization.
\end{itemize}
\label{rmk:2}
\end{remark}
\vspace{1ex}

Based on this profound physical distinction, the FA receiver not only more accurately reflects biological ligand-receptor mechanisms \cite{schulten2000lectures}, but it also fundamentally guarantees the mathematical robustness of optimal waveform designs. Consequently, to ensure a reliable link control framework, our subsequent optimization will focus exclusively on the FA spherical receiver.

\section{Detection Rule and BEP Analysis}
\label{sec:BEP_analysis}

This section evaluates the end-to-end system performance using BEP as the primary metric. To systematically analyze the impact of the time-varying electric field, we adopt a MAP detection rule combined with DFE. For analytical tractability, we assume a perfect decision-feedback implementation where all previous bits $\mathbf{b}_{[i-1]} = [b_1, \dots, b_{i-1}]$ are correctly decoded and available at the receiver.

Recalling the statistical framework from our previous work \cite{Chou2022}, the MAP detection rule is formulated as
\begin{IEEEeqnarray}{rCl}
    \frac{p(y[i] \mid b_i = 1, \mathbf{b}_{[i-1]})}{p(y[i] \mid b_i = 0, \mathbf{b}_{[i-1]})} \mathop{\gtrless}_{H_0}^{H_1} \frac{p(b_i = 0)}{p(b_i = 1)}, \label{eq:map_rule}
\end{IEEEeqnarray}
where hypotheses $H_1$ and $H_0$ correspond to the transmission of bit `1' and `0', respectively. 

To characterize the detection statistics, we introduce the superscript $[i, j]$ for the $i$-th symbol conditioned on the $j$-th candidate sequence of preceding bits. As established in \cite{Chou2022}, let $\mu_0^{[i, j]}$ be the expected received molecule count under $H_0$, which accounts for the residual ISI from all previous emissions, we have
\begin{IEEEeqnarray}{rCl}
    \mu_0^{[i, j]} = N \sum_{k=2}^{i-1} b_{i-k}^{[j]} p_{\mathrm{R}}[k]. \label{eq:mean_h0}
\end{IEEEeqnarray}
Under $H_1$, the expected count includes the current signal component, such that $\mu_1^{[i, j]} = N p_{\mathrm{R}}[1] + \mu_0^{[i, j]}$. The corresponding conditional variances are given by $\sigma_0^{[i, j]} = ( \sum_{k=2}^{i-1} b_{i-k}^{[j]} \sigma_k^2 )^{1/2}$ and $\sigma_1^{[i, j]} = ( \sigma_1^2 + (\sigma_0^{[i, j]})^2 )^{1/2}$, where $\sigma_k^2 = N p_{\mathrm{R}}[k](1 - p_{\mathrm{R}}[k])$ represents the signal-dependent counting noise.

Based on the Gaussian approximation, the observation $y[i]$ follows the conditional distributions $\mathcal{N}(\mu_1^{[i, j]}, (\sigma_1^{[i, j]})^2)$ and $\mathcal{N}(\mu_0^{[i, j]}, (\sigma_0^{[i, j]})^2)$. By determining the optimal MAP decision threshold $\gamma^{[i, j]}$, we recall from \cite{Chou2022} that the conditional false alarm probability $P_{\mathrm{FA}}^{[i, j]}$ and detection probability $P_{\mathrm{D}}^{[i, j]}$ are formulated as
\begin{IEEEeqnarray}{rCl}
    P_{\mathrm{FA}}^{[i, j]} &=& Q\left( \frac{\gamma^{[i, j]} - \mu_0^{[i, j]}}{\sigma_0^{[i, j]}} \right), \nonumber \\
    P_{\mathrm{D}}^{[i, j]} &=& Q\left( \frac{\gamma^{[i, j]} - \mu_1^{[i, j]}}{\sigma_1^{[i, j]}} \right), \label{eq:p_fa_p_d}
\end{IEEEeqnarray}
where $Q(\cdot)$ is the standard $Q$-function. Detailed derivations of the exact optimal threshold $\gamma^{[i, j]}$ can be found in \cite{Chou2022}.

The average BEP $P_e^{[i]}$ is then obtained by marginalizing over all $2^{i-1}$ possible ISI sequences, 
\begin{IEEEeqnarray}{rCl}
    P_e^{[i]} &=& \sum_{j=1}^{2^{i-1}} p( \mathbf{b}_{[i-1]}^{[j]} ) \Bigl[ p(b_i=0) P_{\mathrm{FA}}^{[i,j]} \nonumber \\
    && \qquad \qquad \quad + p(b_i=1)(1 - P_{\mathrm{D}}^{[i,j]} ) \Bigr]. \label{eq:pe_isi}
\end{IEEEeqnarray}

\vspace{1ex}
\begin{remark}[Surrogate Objective for System Optimization]
Directly minimizing the average BEP in \eqref{eq:pe_isi} with respect to the time-varying electric field $\mathbf{E}(t)$ is analytically prohibitive due to the highly non-linear dependence of the threshold on the entire channel coefficient sequence $\{p_{\mathrm{R}}[k]\}$. In prior work \cite{Chou2022}, such intractability necessitated computationally expensive exhaustive searches or complex gradient descent approximations (e.g., MinEP and MaxSIR algorithms) to find suboptimal field waveforms.
\end{remark}
\vspace{1ex}

To circumvent this bottleneck in our generalized framework, we analyze the system in an ideal ISI-free regime where $p_{\mathrm{R}}[k] \approx 0$ for $k \ge 2$. In this condition, $P_{\mathrm{FA}}^{[i]} \to 0$ and the detection probability is purely dominated by the target signal,
\begin{IEEEeqnarray}{rCl}
    P_{\mathrm{D}}^{[i]} \approx 1 - Q\left( \sqrt{N p_{\mathrm{R}}[1]} \right). \label{eq:pd_approx_final}
\end{IEEEeqnarray}
Equation \eqref{eq:pd_approx_final} reveals a crucial physical insight: the BEP monotonically decreases as the first-slot received probability $p_{\mathrm{R}}[1]$ increases. While this approximation is universal, its practical utility as an optimization objective strictly depends on the receiver architecture. As highlighted in Remark~\ref{rmk:1}, unlike passive sensing which suffers from an intrinsic physical limitation, the FA receiver guarantees a strictly monotonic and concave response to the directed electric field.

Consequently, for the FA architecture, maximizing $p_{\mathrm{R}}[1]$ serves as a robust, mathematically guaranteed surrogate objective for BEP minimization. Guided by this insight, we bypass the intractable exact BEP formulation and adopt a computationally efficient two-phase sequential strategy: first designing the field waveform to aggressively maximize $p_{\mathrm{R}}[1]$, and subsequently utilizing the residual energy to suppress the leading ISI term $p_{\mathrm{R}}[2]$.

\section{Waveform Optimization and System Design}\label{sec:wf_optimization}

\begin{figure*}[t]
    \centering
    \begin{minipage}[t]{0.48\textwidth}
        \centering
        \begin{tikzpicture}[>=Stealth]

\begin{axis}[
    width=13cm,
    height=8cm,
    xmin=0,xmax=16,
    ymin=-2.4,ymax=1.6,
    axis lines=none,
    clip=false
]

\draw[->,very thick] (axis cs:0,0) -- (axis cs:9,0);
\draw[->,very thick] (axis cs:0,0) -- (axis cs:0,1.5);

\draw[->,very thick] (axis cs:0,-1.5) -- (axis cs:9,-1.5);
\draw[-,very thick] (axis cs:0,-1.5) -- (axis cs:0,0);

\addplot[
thick,
smooth,
blue!50!black,
samples=400,
domain=0:6
]
{20*x^3*exp(-3*x)};

\addplot[
thick,
smooth,
blue!50!black,
samples=400,
domain=3:9
]
{20*(x-3)^3*exp(-3*(x-3))};

\draw[purple,->,dashed,thick] (axis cs:1,-1.5) -- (axis cs:1,1);
\node[purple] at (axis cs:0.9,1.2) {$t_{\text{peak}}$};

\draw[blue,dashed,very thick] (axis cs:3,-1.5) -- (axis cs:3,1.2);
\draw[blue,dashed,very thick] (axis cs:6,-1.5) -- (axis cs:6,1.2);

\node[blue] at (axis cs:3,1.3) {$T_b$};

\draw[red,<->,thick] (axis cs:0,-0.4) -- (axis cs:1,-0.4);
\node[red] at (axis cs:0.5,-0.2) {Phase I};

\draw[brown!80!black,<->,thick] (axis cs:2.3,-0.4) -- (axis cs:3.0,-0.4);
\node[brown!80!black] at (axis cs:2.75,-0.2) {Phase II};

\draw[green!60!black,very thick]
(axis cs:0,-0.8) --          
(axis cs:1.0,-0.8) --        
(axis cs:1.0,-1.5) --        
(axis cs:2.5,-1.5) --        
(axis cs:2.5,-1.9) --        
(axis cs:3.0,-1.9) --        
(axis cs:3.0,-0.8) --        
(axis cs:4.2,-0.8) --
(axis cs:4.2,-1.5) --
(axis cs:5.5,-1.5) --
(axis cs:5.5,-1.9) --        
(axis cs:6.0,-1.9);

\node[green!60!black,left] at (axis cs:0,-0.8) {$V_1$};
\node[green!60!black,left] at (axis cs:0,-1.9) {$V_2$}; 

\draw[brown!80!black,->,thick] (axis cs:2,-1.8) -- (axis cs:2.5,-1.9);
\node[brown!80!black] at (axis cs:1.8,-1.7) {$\beta T_b$};

\node at (axis cs:8,0.5) {$\cdots$};
\node at (axis cs:8,-0.8) {$\cdots$};

\node at (axis cs:5,1.6) {Hitting probability density $f_{\mathrm{hit}}^{\Phi}(t)$};
\node at (axis cs:5,-2.3) {Assisted electric-field waveform $E_1(t)$};

\end{axis}

\end{tikzpicture}
        \medskip
        \centerline{\textbf{(a)} Fully-absorbing (active) spherical Rx}
    \end{minipage}
    \hfill 
    \begin{minipage}[t]{0.48\textwidth}
        \centering
        \begin{tikzpicture}[>=Stealth]

\begin{axis}[
    width=13cm,
    height=8cm,
    xmin=0,xmax=16,
    ymin=-2.4,ymax=1.6,
    axis lines=none,
    clip=false
]

\draw[->,very thick] (axis cs:0,0) -- (axis cs:9,0);
\draw[->,very thick] (axis cs:0,0) -- (axis cs:0,1.5);

\draw[->,very thick] (axis cs:0,-1.5) -- (axis cs:9,-1.5);
\draw[-,very thick] (axis cs:0,-1.5) -- (axis cs:0,0);


\addplot[
thick,
smooth,
blue!50!black,
samples=400,
domain=0:6
]
{20*x^3*exp(-3*x)};

\addplot[
thick,
smooth,
blue!50!black,
samples=400,
domain=3:9
]
{20*(x-3)^3*exp(-3*(x-3))};

\draw[purple,->,dashed,thick] (axis cs:1,-1.5) -- (axis cs:1,1);
\node[purple] at (axis cs:0.9,1.2) {$t_s$};

\draw[blue,dashed,very thick] (axis cs:3,-1.5) -- (axis cs:3,1.2);
\draw[blue,dashed,very thick] (axis cs:6,-1.5) -- (axis cs:6,1.2);

\node[blue] at (axis cs:3,1.3) {$T_b$};

\draw[red,<->,thick] (axis cs:0,-0.4) -- (axis cs:1,-0.4);
\node[red] at (axis cs:0.5,-0.2) {Phase I};

\draw[brown!80!black,<->,thick] (axis cs:1,-0.4) -- (axis cs:3,-0.4);
\node[brown!80!black] at (axis cs:2,-0.6) {Phase II};

\draw[green!60!black,very thick]
(axis cs:0,-0.8) --          
(axis cs:1.0,-0.8) --        
(axis cs:1.0,-1.2) --        
(axis cs:3.0,-1.2) --        
(axis cs:3.0,-0.8) --        
(axis cs:4.0,-0.8) --        
(axis cs:4.0,-1.2) --        
(axis cs:6.0,-1.2);          

\node[green!60!black,left] at (axis cs:0,-0.8) {$V_1$};
\node[green!60!black,left] at (axis cs:0,-1.2) {$V_2$};

\node at (axis cs:8,0.5) {$\cdots$};
\node at (axis cs:8,-0.8) {$\cdots$};

\node at (axis cs:5,1.6) {Sensing probability $p_{\mathrm{s}}(t)$};
\node at (axis cs:5,-1.8) {Assisted electric-field waveform $E_1(t)$};

\end{axis}

\end{tikzpicture}
        \medskip
        \centerline{\textbf{(b)} Passive spherical Rx}
    \end{minipage}

    \caption{
        Illustration of the optimized assisted electric-field waveforms $E_1(t)$ for the two receiver architectures. 
        \textbf{(a)} \textbf{Fully-absorbing Rx:} The waveform is designed to maximize the hitting density at $t_{\mathrm{peak}}$ through the MHP method, with a strategic suppression phase $V_2$ to divert residual particles. 
        \textbf{(b)} \textbf{Passive Rx:} The waveform optimizes the sensing probability at the sampling time $t_s$ using the MSP method. 
        Both designs are generated by the unified MRP engine under a total energy constraint $\xi$.
    }
    \label{fig:AE_waveforms}
\end{figure*}
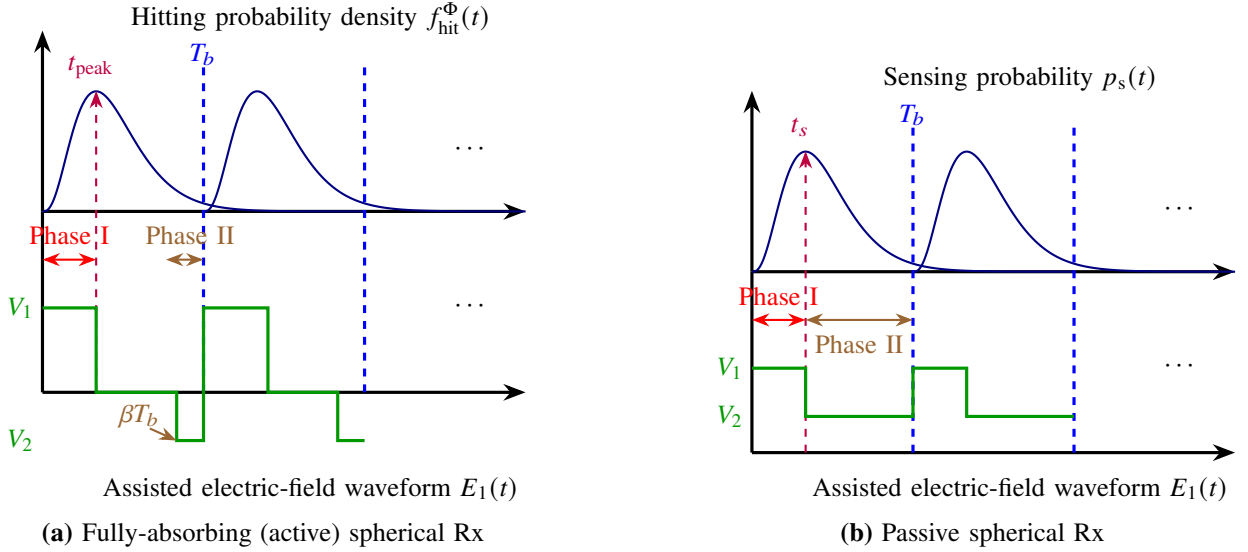

Driven by the analytical insight from Section~\ref{sec:BEP_analysis} that the first-slot received probability $p_{R}[1]$ dictates the system BEP, we now formulate the assisted electric field design as a tractable optimization problem. In alignment with the robustness rationale established in Remark~\ref{rmk:2}, this section exclusively focuses on the FA spherical receiver. The general structure of the optimized assisted electric-field waveforms for both receiver architectures is illustrated in Fig.~\ref{fig:AE_waveforms}.

\subsection{Problem Formulation and Energy Constraint}

To actively shape the channel impulse response, we design a periodic, time-varying assisted electric field $\mathbf{E}(t)$. For design tractability and to maximize transport efficiency along the transmission link, we orient the composite field along the $x_1$-axis, coinciding with the Tx--Rx trajectory. This allows us to simplify the design variable to a scalar electric field component $E_1(t)$.

The molecular propagation is simultaneously influenced by a time-varying background flow $u_1(t)$ acting along the Tx--Rx axis. The background flow is selected as in \cite{koikeakino2017molecular},
\begin{IEEEeqnarray}{rCl}
    u_1(t) &=& \mu F_0 \cos^2 \left( \frac{\pi t}{T_d} \right), \label{eq:drift_field}
\end{IEEEeqnarray}
where $\mu$ is the mechanical mobility, $F_0$ is the peak flow force amplitude, and $T_d$ is the flow period. The transmitter is located at $\mathbf{x}_0 = [x_0, 0, 0]$ where $x_0 = -d_0$. For analytical simplicity and without loss of generality, the initial phase of the background flow is set to zero.

The system operates under a fundamental energy constraint $\xi$, which bounds the kinetic energy imparted by the external field over a symbol period $T_b$. The normalized energy constraint is defined as
\begin{IEEEeqnarray}{rCl}
    \int_{0}^{T_b} |E_1(\tau)|^2 d\tau \leq \xi. \label{eq:energy_constraint}
\end{IEEEeqnarray}
We decouple the waveform $E_1(t)$ into two sequential phases: a signal-enhancing field $E_1^{\mathrm{S}}(t)$ to maximize $p_{\mathrm{R}}[1]$, followed by an ISI-suppressing field $E_1^{\mathrm{R}}(t)$ utilizing the residual energy $\xi_{\mathrm{res}}$.

\subsection{Phase I: Signal Enhancement via MHP Approximation}

For the active receiver, maximizing $p_{\mathrm{R}}[1] = \int_0^{T_b} f_{\mathrm{hit}}^{\Phi}(t) dt$ is computationally prohibitive due to the complex temporal integral of Bessel functions. To establish a tractable surrogate, we maximize the instantaneous hitting density at the peak-receiving location $\mathbf{x}_{\mathrm{peak}} = [-r_{\mathrm{Rx}}, 0, 0]$ at a target peak time $t_{\mathrm{peak}}$. Based on the universal engine in \eqref{eq:f_trans_pdf_final}, the control variable $E_1^{\mathrm{S}}(t)$ influences the hitting density through the exponential Girsanov weight, 
\begin{IEEEeqnarray}{rCl}
    \Psi(T) &=& \exp\left( -\frac{T}{4D} \left\| \boldsymbol{\Phi}_{\mathrm{eff}}(T) - \frac{\mathbf{x}_{\mathrm{peak}} - \mathbf{x}_0}{T} \right\|^2 + C_{\mathbf{x}} \right), \quad \quad
\end{IEEEeqnarray}
where $C_{\mathbf{x}}$ is a constant independent of the electric field. This reformulates the maximization into a \textit{distance-minimization problem}:
\begin{IEEEeqnarray}{rl}
    (\mathrm{P}_{\mathrm{FA}}) \quad \argmin_{E_1^{\mathrm{S}}} & \quad \left\| \boldsymbol{\Phi}_{\mathrm{eff}}(t_{\mathrm{peak}}) - \frac{ \mathbf{x}_{\mathrm{peak}} - \mathbf{x}_0 }{ t_{\mathrm{peak}} } \right\|^2. \label{eq:ab_approx_opt_4}
\end{IEEEeqnarray}
The global minimum is achieved when the effective drift completely bridges the spatial gap within the peak time. Solving this condition yields the ideal unconstrained field velocity $q_{\mathrm{S}}^{\mathrm{A}}$. If the energy required exceeds the budget $\xi$, the waveform is truncated to $V_1^* = \mathrm{sgn}(q_{\mathrm{S}}^{\mathrm{A}}) \sqrt{\xi/t_{\mathrm{peak}}}$.

\subsection{Phase II: ISI Suppression and the MRP Algorithm}

The residual energy $\xi_{\mathrm{res}} = \xi - (V_1^*)^2 t_{\mathrm{peak}}$ is deployed to suppress $p_{\mathrm{R}}[2]$. Since the FA receiver continuously integrates arrivals, we deploy an opposing drift field late in the symbol period to deviate residual particles. We define a fraction $\beta \in [0,1]$ for the suppression onset, applying a constant field $V_2$ during $t \in [\beta T_b, T_b]$. 

The velocity allocation is consolidated into the \textit{Maximize Received Probability} algorithm, presented in Algorithm~\ref{alg:maxRP}, where the temporal durations for the fully-absorbing architecture map to $T_{p1} = t_{\mathrm{peak}}$ and $T_{p2} = (1-\beta)T_b$. The final composite electric field is compactly expressed as
\begin{IEEEeqnarray}{rCl}
    E_1(t) &=&
    \begin{cases}
        V_1^*, & (i-1)T_b \le t < t_{\mathrm{peak}} + (i-1)T_b, \\
        V_2^*, & \beta T_b + (i-1)T_b \le t < iT_b, \\
        0, & \text{otherwise}.
    \end{cases} \label{eq:E1_absorbing}
\end{IEEEeqnarray}

\begin{algorithm}[t]
  \caption{Maximize Received Probability (MRP) Engine}
  \label{alg:maxRP}
  \begin{algorithmic}[1]
    \Require Target displacements $q_{\mathrm{S}}, q_{\mathrm{R}}$; durations $T_{p1}, T_{p2}$; budget $\xi$ 
    \Ensure Optimal electric field magnitudes $(V_1^*, V_2^*)$ 
    \If{$|q_{\mathrm{S}}| \geq \sqrt{ \xi / T_{p1} }$} \Comment{Insufficient energy to reach optimal peak} 
      \State $V_1^* \gets \mathrm{sgn}(q_{\mathrm{S}}) \cdot \sqrt{ \xi / T_{p1} }$ 
      \State $V_2^* \gets 0$ \Comment{No residual energy for ISI suppression} 
    \Else \Comment{Unconstrained optimum achieved} 
      \State $V_1^* \gets q_{\mathrm{S}}$ 
      \If{$T_{p1} = T_b$} 
        \State $V_2^* \gets 0$ 
      \Else 
        \State $\xi_{\mathrm{res}} \gets \xi - (V_1^*)^2 \cdot T_{p1}$ \Comment{Compute residual energy} 
        \State $V_2^* \gets -\mathrm{sgn}(q_{\mathrm{R}}) \cdot \sqrt{ \xi_{\mathrm{res}} / T_{p2} }$ \Comment{Apply maximum suppression} 
      \EndIf 
    \EndIf 
  \end{algorithmic}
\end{algorithm}

\vspace{1ex}
\begin{remark}[Universality of the MRP Engine]
While we focus on the FA receiver (i.e., MHP), the logic in Algorithm~\ref{alg:maxRP} is inherently universal. By substituting $\mathbf{x}_{\mathrm{peak}}$ with the volumetric center, the engine generates the \textit{Maximize Sensing Probability} solution for PA receivers (see Appendix~\ref{appendix:MSP_design}), underscoring the cohesiveness of our framework.
\end{remark}
\vspace{1ex}

\section{Numerical Validation and Performance Evaluation}\label{sec:numerical}

In this section, we provide a comprehensive numerical evaluation of the proposed FAMC framework. To ensure the full reproducibility of our results, we first detail the physical parameters and simulation environment. Subsequently, we validate the accuracy of the Girsanov-based analytical CIR expressions via particle-based Monte Carlo simulations. Finally, we evaluate the system-level performance, demonstrating the efficacy of the proposed MRP waveform optimization algorithms in terms of signal-to-ISI enhancement and BEP reduction.

\subsection{Simulation Setup}

Unless otherwise specified, the system parameters for the numerical evaluations are configured as follows. The spherical receiver has a radius of $r_{\mathrm{Rx}} = 10~\mu\mathrm{m}$, and the distance from the point transmitter to the receiver center is $d_0 = 30~\mu\mathrm{m}$. The message particles are calcium ions ($\mathrm{Ca}^{2+}$) with a valence of $z = 2$ and a diffusion coefficient of $D = 7.5 \times 10^{-6}~\mathrm{cm^2/s}$ in a fluidic environment at an absolute temperature of $T = 300~\mathrm{K}$ \cite{koikeakino2017molecular}. For the passive receiver, the sampling time is $t_s = 0.1~\mathrm{s}$, whereas for the fully-absorbing receiver, the target peak time is set to $t_{\mathrm{peak}} = 0.1~\mathrm{s}$. The communication symbol duration is strictly defined as $T_b = 2~\mathrm{s}$. To explicitly evaluate the impact of the ISI suppression timing in the proposed MHP method, we investigate two suppression onset fractions: $\beta = 0.5$ and $\beta = 0.8$. Furthermore, to accurately capture the ISI dynamics, the BEP evaluation accounts for the residual molecules from the preceding two symbol intervals (i.e., evaluating up to the third symbol). The background flow is characterized by a drift duration of $T_{\mathrm{d}} = 1~\mathrm{s}$, a mobility factor of $\mu = 1.77 \times 10^{11}~\mathrm{s/kg}$, and a force amplitude of $F_0 = 4.18 \times 10^{-15}~\mathrm{N}$, which align with typical micro-fluidic modeling \cite{koikeakino2017molecular}.

\begin{figure}[t]
    \begin{minipage}[t]{0.48\textwidth}
    \centering
    \pgfplotsset{compat=1.18} 

\begin{tikzpicture}
\begin{axis}[
    width=\textwidth,
    height=6cm,
    xlabel={Sampling Time (sec.)},
    ylabel={Amplitude (cm/s)},
    label style={font=\small},
    xmin=0, xmax=2,
    ymin=0, ymax=1.5e-3,
    xtick={0,0.5,1,1.5,2},
    ytick={0,0.5e-3,1e-3,1.5e-3},
    yticklabels={0,0.5,1,1.5},
    tick label style={font=\small},
    scaled y ticks=false,
    legend style={font=\small,at={(0.98,0.98)},anchor=north east},
    grid=major
]

\addplot[
    blue,
    thick,
    mark=diamond,
    mark size=3pt,
    const plot
]
coordinates {
    (0,0.417e-3)
    (0.5,0.7203e-3)
    (1,0.0001e-3)
    (1.5,0.3023e-3)
    (2,0.3023e-3)
};
\addlegendentry{Waveform field on $x_1$-axis}

\addplot[
    red,
    thick,
    mark=o,
    mark size=3pt,
    const plot
]
coordinates {
    (0,0.1468e-3)
    (0.5,0.0923e-3)
    (1,0.1863e-3)
    (1.5,0.3456e-3)
    (2,0.3456e-3)
};
\addlegendentry{Waveform field on $x_2$-axis}

\addplot[
    orange!80!black,
    thick,
    mark=square,
    mark size=3pt,
    const plot
]
coordinates {
    (0,0.3968e-3)
    (0.5,0.5388e-3)
    (1,0.4192e-3)
    (1.5,0.6852e-3)
    (2,0.6852e-3)
};
\addlegendentry{Waveform field on $x_3$-axis}

\node at (axis description cs:0.02,0.98) [anchor=north west] {$\times 10^{-3}$};

\end{axis}
\end{tikzpicture}
    \end{minipage}
    \caption{Arbitrary time-varying drift realizations $\boldsymbol{\Phi}(t)$.}
    \label{fig:arbitrary_time_varying_drift}
\end{figure}
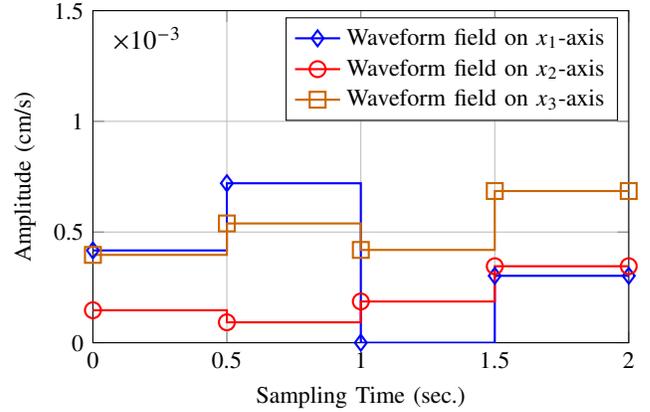

\subsection{CIR Verification with Particle-Based Simulation}

To validate the accuracy of the proposed Girsanov-based analytical CIR expressions, we compare the theoretical distributions against particle-based Monte Carlo simulations tracing three-dimensional Brownian random walks \cite{berg1993random}. To ensure high statistical fidelity, the Monte Carlo simulations independently track $N = 10^6$ message particles for each scenario. Furthermore, we construct arbitrary time-varying composite drift fields $\boldsymbol{\Phi}(t)$, generated as 3-D piecewise constant vector fields, and observe the reception statistics for both the active and passive receiver architectures. For exact reproducibility and a consistent comparison between the two architectures, the drift field realization shown in Fig.~\ref{fig:arbitrary_time_varying_drift} is generated by initializing the pseudo-random number generator with a fixed seed (i.e., seed = 1).

\begin{figure*}[t]
    \centering
    \includegraphics[width=0.48\textwidth,height=5.2cm]{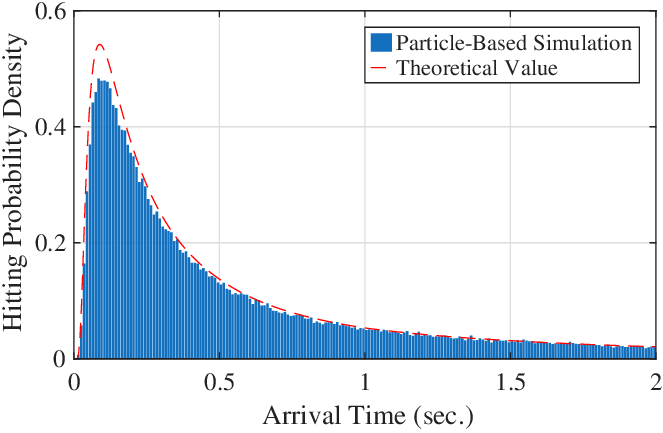}
    \hfill
    \includegraphics[width=0.48\textwidth,height=5.2cm]{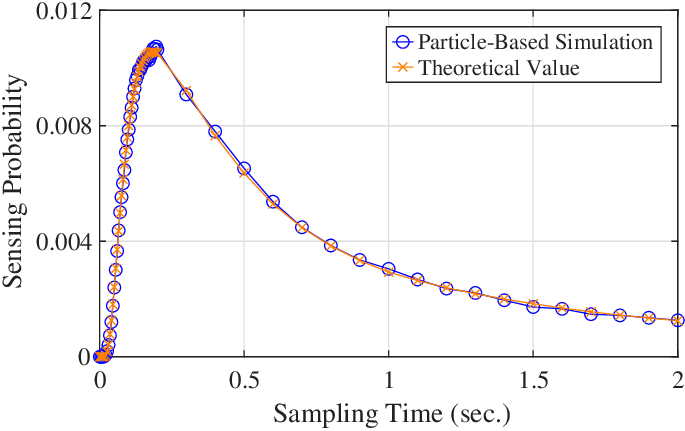}
    \caption{
        Verification of the analytical CIR for the \textit{fully-absorbing spherical receiver} and the \textit{passive spherical receiver}. 
        The left panel compares the analytical hitting probability density (dashed lines, computed via \eqref{eq:hit_time_varying_final}) with particle-based Monte Carlo simulations. The right panel compares the analytical sensing probability (dashed lines, computed via \eqref{eq:spherical_rx}) with particle-based Monte Carlo simulations.   
    }
    \label{fig:verification}
\end{figure*}

As illustrated in the left and right panels of Fig.~\ref{fig:verification}, the hitting probability density in \eqref{eq:hit_time_varying_final} for the fully-absorbing receiver and the analytical sensing probability in \eqref{eq:spherical_rx} for the passive receiver perfectly track the simulated particle behaviors. This excellent agreement confirms that the EDA successfully captures the path-dependent weighting induced by the time-varying electric field, thereby preserving the mathematical tractability required for subsequent system optimization.

\subsection{Impact of Energy Constraints and Waveform Design}

Subsequently, we investigate how the optimal allocation of the assisted electric field affects the system performance under varying energy budgets $\xi$. Specifically, we evaluate the signal-to-ISI difference, defined as $p_{\mathrm{R}}[1] - p_{\mathrm{R}}[2]$, alongside the overall BEP for energy constraints up to $\xi = 200~\mathrm{V^2 \cdot s/m^2}$. The proposed MHP and MSP methods are executed via the unified MRP engine (Algorithm~\ref{alg:maxRP}), utilizing the analytical parameters for the passive architecture derived in Appendix~\ref{appendix:MSP_design}. These proposed designs are systematically compared against a pure diffusion baseline (i.e., No Drift) and a constant-field approach (i.e., Undesign) that consumes the energy budget evenly over the symbol duration $T_b$.

\begin{figure*}[t]
    \centering
    \includegraphics[width=0.48\textwidth,height=5.0cm]{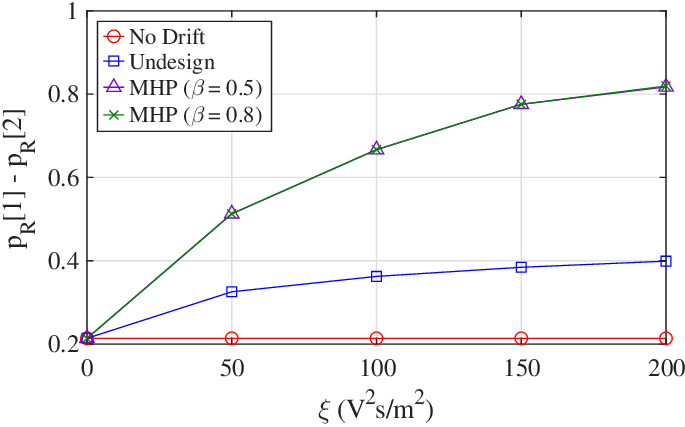}
    \hfill
    \includegraphics[width=0.48\textwidth,height=5.0cm]{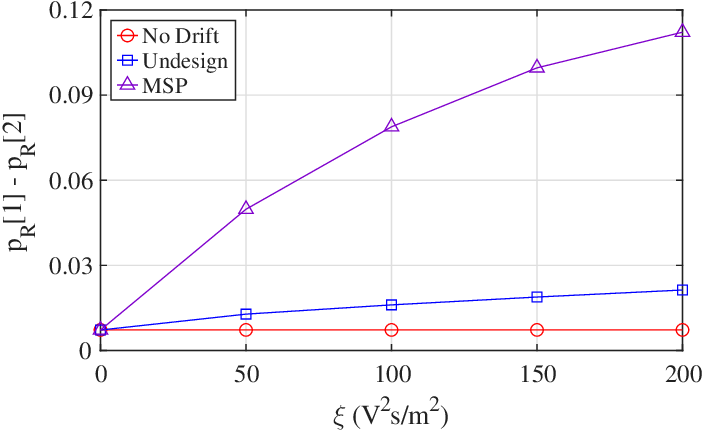}
    \\[2ex]
    \includegraphics[width=0.48\textwidth,height=5.0cm]{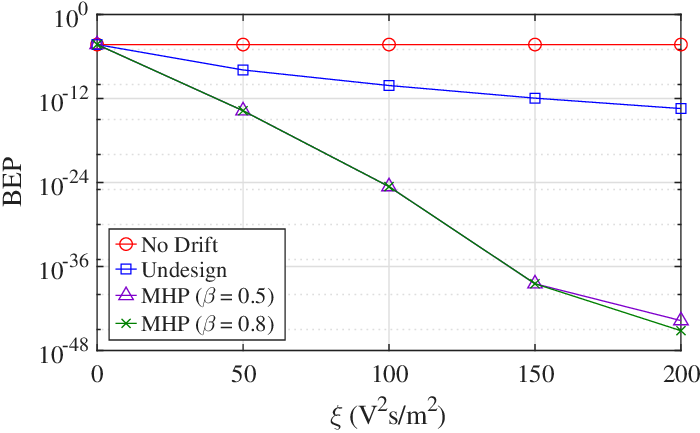}
    \hfill
    \includegraphics[width=0.48\textwidth,height=5.0cm]{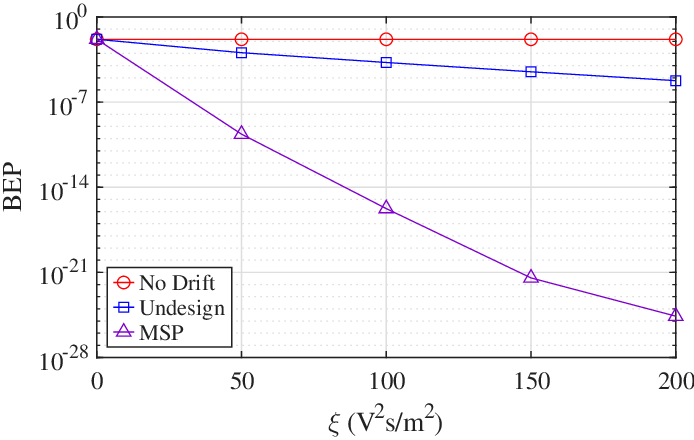}
    \caption{
        Performance metrics versus energy constraint $\xi$ under different electric-field design methods. \textbf{(Top Row)} The signal-to-ISI difference $p_{\mathrm{R}}[1] - p_{\mathrm{R}}[2]$ for FA (left) and PA (right) receivers. \textbf{(Bottom Row)} The corresponding BEP for FA (left) and PA (right) receivers. The fully-absorbing architecture (left) guarantees robust optimization and monotonic improvement, whereas the passive receiver (right) faces performance saturation at higher energy levels due to its intrinsic physical limitation.
    }
    \label{fig:PR_Xi_compare}
\end{figure*}

The numerical results presented in Fig.~\ref{fig:PR_Xi_compare} provide a profound physical validation of the robustness rationale established in Remark 2. For the fully-absorbing receiver (left panels), the MHP method exhibits strict monotonicity. As $\xi$ increases, the algorithm efficiently allocates energy to accelerate particles, steadily maximizing the target signal $p_{\mathrm{R}}[1]$ while effectively utilizing the residual energy to deviate trailing particles, thereby suppressing $p_{\mathrm{R}}[2]$. Consequently, the signal-to-ISI difference expands continuously, leading to a rapid and robust decay in BEP. Furthermore, comparing the two onset fractions, a delayed suppression phase ($\beta=0.8$) yields superior ISI mitigation and a strictly lower BEP compared to an earlier onset ($\beta=0.5$) when sufficient energy is available.

Conversely, the passive receiver (right panels) is fundamentally limited by its snapshot sensing mechanism. Although the signal-to-ISI difference continues to improve as $\xi$ increases towards $200~\mathrm{V^2 \cdot s/m^2}$, the absolute magnitude of this gap remains severely bounded compared to the active architecture. Specifically, the passive receiver only achieves a maximum difference of approximately $0.11$, whereas the fully-absorbing receiver rapidly approaches $0.8$. Because the powerful electric field forces the molecular cloud to traverse the sensing volume rapidly, the peak concentration captured at any single sampling instant is fundamentally lower than the cumulative number of molecules absorbed over a time interval. This intrinsic physical limitation inherently bottlenecks the overall BEP reduction, preventing the passive receiver from reaching the ultra-reliable performance levels of the active architecture. This stark contrast confirms that the cumulative absorption mechanism fundamentally guarantees superior optimization potential, validating our primary design focus.


\subsection{BEP Performance Evaluation}

\begin{figure*}[t]
    \centering
    \includegraphics[width=0.48\textwidth,height=5.0cm]{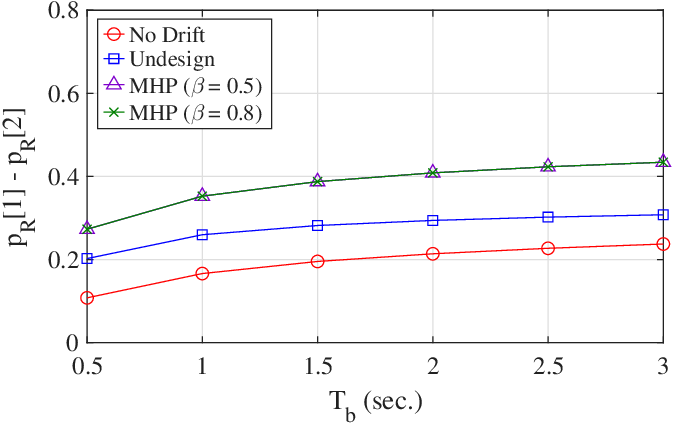}
    \hfill
    \includegraphics[width=0.48\textwidth,height=5.0cm]{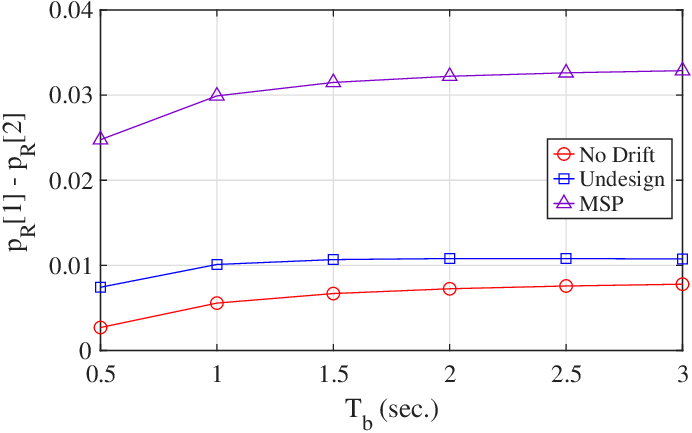}
    \\[1ex]
    \includegraphics[width=0.48\textwidth,height=5.0cm]{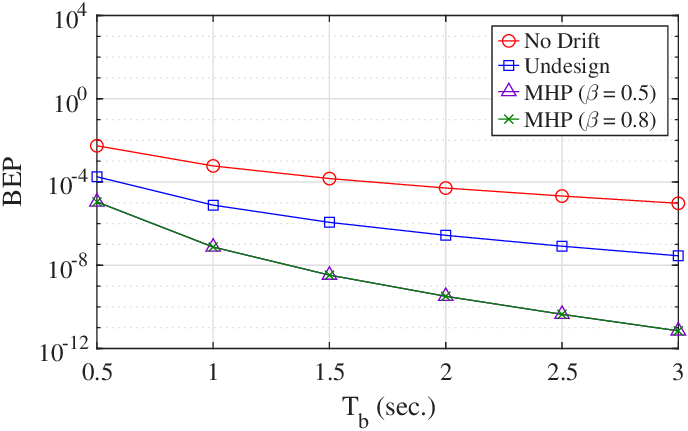}
    \hfill
    \includegraphics[width=0.48\textwidth,height=5.0cm]{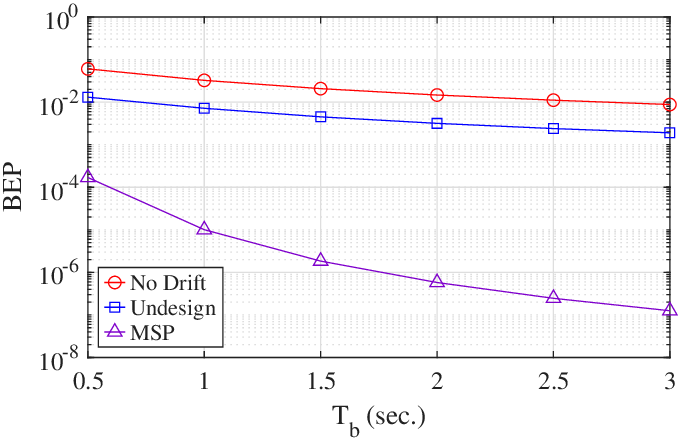}
    \caption{
        Performance metrics versus symbol duration $T_b$ under an energy budget of $\xi=25~\mathrm{V^2 \cdot s/m^2}$. \textbf{(Top Row)} The signal-to-ISI difference $p_{\mathrm{R}}[1] - p_{\mathrm{R}}[2]$ for FA (left) and PA (right) receivers. \textbf{(Bottom Row)} The corresponding BEP for FA (left) and PA (right) receivers. The proposed MHP and MSP methods significantly and consistently outperform the Undesign (constant-field) baseline across all evaluated symbol durations.
    }
    \label{fig:BEP_Tb_compare}
\end{figure*}

Finally, we evaluate the overall system reliability across different symbol durations $T_b$. Fig.~\ref{fig:BEP_Tb_compare} presents both the signal-to-ISI difference ($p_{\mathrm{R}}[1] - p_{\mathrm{R}}[2]$) and the BEP under a moderate energy budget of $\xi = 25~\mathrm{V^2 \cdot s/m^2}$. The transmission involves $N = 100$ and $N = 1000$ particles for the absorbing and passive receivers, respectively.

As observed in the top panels of Fig.~\ref{fig:BEP_Tb_compare}, extending the symbol duration $T_b$ provides an expanded temporal window for the electric field to actively shape the molecular transport, thereby increasing the separation between the target signal and the leading ISI. This direct improvement in the physical channel response translates into a significant reduction in BEP, as demonstrated in the bottom panels.

The proposed two-phase waveform designs (MHP and MSP) consistently outperform both the pure diffusion (No Drift) and the constant-field (Undesign) baselines across all evaluated symbol durations. For the active receiver operating under the MHP strategy, employing a delayed suppression onset ($\beta=0.8$) yields a more robust signal-to-ISI gap and a lower BEP compared to an earlier onset ($\beta=0.5$), further highlighting the critical importance of timely ISI mitigation.
By efficiently allocating the constrained energy budget, the lightweight MRP engine guarantees highly reliable detection performance, establishing it as a highly practical solution for low-complexity bio-nanoreceivers.

\section{Conclusion}\label{sec:conclude}

This paper has established a comprehensive analytical and optimization framework for molecular communication systems operating under time-varying composite drift fields. 
By harnessing the CMG theorem, we successfully decoupled the temporal variations of the assistive electric field from the complex spatial boundaries, deriving analytically tractable CIRs via effective drift mapping for FA receivers, and exact CIR expressions for PA spherical receivers. These analytical models were rigorously validated via particle-based simulations, confirming that the EDA captures the path-dependent weighting of dynamic fields with high fidelity.

Building upon this foundation, we identified the first-slot received probability as the primary determinant of BEP performance in decision-feedback detection architectures. Recognizing the intrinsic physical limitations of passive sensing, we strategically focused our optimization on the FA receiver, which exhibits a robust monotonic response to assistive fields. We proposed the \textit{Maximize Received Probability} algorithm, which intelligently shapes the electric field into a signal-enhancing acceleration phase and an ISI-suppressing counter-drift phase. Numerical evaluations demonstrate that our design achieves near-optimal performance under strict energy constraints with low computational complexity, offering a scalable and reliable control paradigm for advanced field-assisted bio-nanonetworks.

\appendices

\section{Effective Drift Approximation Based on Measure-Change Framework}
\label{appendix:EDA_approx}

The fundamental framework for evaluating reception statistics via the measure-change framework was established in our previous work \cite{Lee2025Exact3D} for channels with a constant uniform drift. To rigorously justify the EDA utilized in \eqref{eq:radon_nikodym_event_approx} for \emph{time-varying} fields, we extend this Girsanov-based approach. 

We consider the event of interest defined as an arrival within a small spatial-temporal window,
\begin{IEEEeqnarray}{rCl}
    A &:=& \bigl\{t \in [T, T + \Delta T],\; \mathbf{X}_t \in [\mathbf{x}, \mathbf{x} + \Delta \mathbf{x}],\; T < \infty \bigr\}. \quad
\end{IEEEeqnarray}

Recalling the application of the Radon--Nikodym theorem from \cite{Lee2025Exact3D}, the likelihood ratio $M_t$ transforms the zero-drift reference measure $\mathbb{P}$ to the drifted physical measure $\mathbb{Q}$. The probability of event $A$ under the true time-varying drift is evaluated by taking the expectation under the pure diffusion measure,
\begin{IEEEeqnarray}{rCl}
    \mathbb{Q}(A) &=& \mathbb{E}^{\mathbb{P}}[M_t \cdot \mathbbm{1}_A] = \int_\Omega M_t \cdot \mathbbm{1}_A\, d\mathbb{P} \nonumber \\
    &=& \int_\Omega \exp \biggl( \int_0^t \frac{\boldsymbol{\Phi}(\alpha)}{\sqrt{2D}} \cdot d\mathbf{B}_\alpha \nonumber \\
    && \qquad - \frac{1}{2} \int_0^t \frac{\|\boldsymbol{\Phi}(\alpha)\|^2}{2D}\, d\alpha \biggr) \cdot \mathbbm{1}_A\, d\mathbb{P}. \quad \quad \label{eq:rn_exact}
\end{IEEEeqnarray}

Unlike the constant drift scenario in \cite{Lee2025Exact3D} where the stochastic integral simplifies exactly, the arbitrary time-varying field $\boldsymbol{\Phi}(\alpha)$ renders the It\^o integral analytically intractable. To resolve this, we approximate the composite drift by its time-average over the interval $[0,t]$, denoted as $\boldsymbol{\Phi}_{\mathrm{eff}}(t) := \frac{1}{t} \int_0^t \boldsymbol{\Phi}(\alpha)\, d\alpha$. Substituting this effective drift allows us to extract it from the stochastic integral,
\begin{IEEEeqnarray}{rCl}
    \int_0^t \frac{\boldsymbol{\Phi}(\alpha)}{\sqrt{2D}} \cdot d\mathbf{B}_\alpha &\approx& \frac{\boldsymbol{\Phi}_{\mathrm{eff}}(t)}{\sqrt{2D}} \cdot \int_0^t d\mathbf{B}_\alpha \nonumber \\
    &=& \frac{\boldsymbol{\Phi}_{\mathrm{eff}}(t)}{\sqrt{2D}} \cdot \mathbf{B}_t.
\end{IEEEeqnarray}

Following the Brownian motion substitution $\mathbf{B}_t = \frac{\mathbf{X}_t - \mathbf{x}_0}{\sqrt{2D}}$ under the pure diffusion reference measure $\mathbb{P}$, we substitute this dynamic relation back into the simplified Radon--Nikodym derivative to yield
\begin{equation}
    M_t \approx \exp \biggl( \frac{\boldsymbol{\Phi}_{\mathrm{eff}}(t) \cdot (\mathbf{X}_t - \mathbf{x}_0)}{2D} - \frac{\|\boldsymbol{\Phi}_{\mathrm{eff}}(t)\|^2 t}{4D} \biggr). \label{eq:rn_expand}
\end{equation}

Finally, the indicator function $\mathbbm{1}_A$ tightly restricts the spatial and temporal variables to $\mathbf{X}_t \approx \mathbf{x}$ and $t \approx T$. Consequently, the exponential term becomes a deterministic weighting factor that can be factored out of the integral,
\begin{equation}
    \mathbb{Q}(A) \approx \exp \biggl( \frac{\boldsymbol{\Phi}_{\mathrm{eff}}(T) \cdot (\mathbf{x} - \mathbf{x}_0)}{2D} - \frac{\|\boldsymbol{\Phi}_{\mathrm{eff}}(T)\|^2 T}{4D} \biggr) \times \mathbb{P}(A). \label{eq:rn_final}
\end{equation}

This explicit formulation confirms that under the EDA, the time-varying effective drift acts as a geometric energy weight applied to the un-drifted probability $\mathbb{P}(A)$. Strictly speaking, approximating the true It\^o integral with the effective drift product introduces a slight second-order variance mismatch, since the variance of the true stochastic integral $\int_0^T \|\boldsymbol{\Phi}(\alpha)\|^2 d\alpha$ differs from that of the approximation $\|\boldsymbol{\Phi}_{\mathrm{eff}}(T)\|^2 T$. However, this difference primarily affects higher-order fluctuation statistics. The EDA should be interpreted as a path-averaged drift approximation that becomes accurate when the drift variation over the hitting time scale is moderate, meaning the first-order geometric weighting definitively dominates the path measure change. This is empirically validated in Fig.~\ref{fig:verification}. Importantly, the approximation preserves the exponential tilting structure induced by the Girsanov transformation. This directly generalizes the static analytical mapping established in \cite{Lee2025Exact3D} to dynamic environments, thereby validating the universal mathematical engine employed in Section~\ref{sec:system_model}.

\section{Derivation of the Sensing Probability for Passive Spherical Receivers}
\label{sec:DSPforPass}

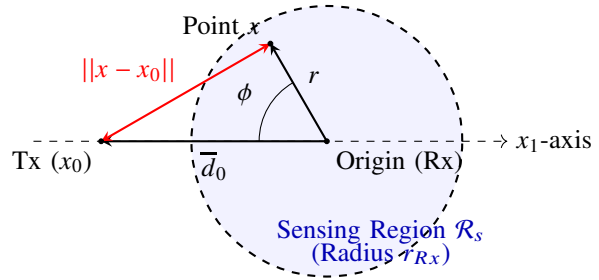
\begin{figure}[htbp]
    \centering
    \begin{tikzpicture}[scale=0.6]
        \coordinate (O) at (0,0);
        \coordinate (Tx) at (-5,0);
        \coordinate (X) at (120:2.5); 
        
        \draw[thick, dashed, fill=blue!5] (O) circle (3);
        \node[text=blue!70!black] at (1.2, -2) {Sensing Region $\mathcal{R}_s$};
        \node[text=blue!70!black] at (1.2, -2.5) {(Radius $r_{Rx}$)};
        
        \draw[->, dashed] (-6.5, 0) -- (4, 0) node[right] {$x_1$-axis};
        
        \draw[->, thick, >=stealth] (O) -- (Tx) node[midway, below] {$\overline{d}_0$};
        \draw[->, thick, >=stealth] (O) -- (X) node[midway, above right] {$r$};
        \draw[<->, thick, >=stealth, red] (X) -- (Tx) node[midway, above left] {$||x - x_0||$};
        
        \filldraw (O) circle (1.5pt) node[below right] {Origin (Rx)};
        \filldraw (Tx) circle (1.5pt) node[below left] {Tx ($x_0$)};
        \filldraw (X) circle (1.5pt) node[above left] {Point $x$};
        
        \pic [draw, "$\phi$", angle eccentricity=1.4, angle radius=0.9cm] {angle = X--O--Tx};
    \end{tikzpicture}
    \caption{Geometric illustration of the PA spherical receiver relative to the point Tx. To align with the system model in Fig.~\ref{fig:system_model}, the Tx is positioned on the negative $x_1$-axis. To facilitate the volume integration over the sensing region $\mathcal{R}_s$, the coordinate origin is set at the center of the receiver. By the law of cosines, the squared distance from an arbitrary point $x$ to the Tx at $x_0$ is evaluated as $||x-x_0||^2 = r^2 + \overline{d}_0^2 - 2r\overline{d}_0 \cos\phi$, which establishes the spatial relationship utilized for the derivation in Appendix~\ref{sec:DSPforPass}.}
    \label{fig:passive_rx_geometry}
\end{figure}

This appendix provides a rigorous derivation of the sensing probability $p_s(T)$ for a PA spherical receiver in a 3-D environment under the reference measure $P$ (zero-drift case). The geometric relationship between the point Tx and the sensing volume, utilized for the subsequent volumetric integration, is illustrated in Fig.~\ref{fig:passive_rx_geometry}. It is worth noting that while the exact expected number of molecules for a spherical receiver was previously formulated by Noel \textit{et al.} \cite[Appendix]{noel2013using} in a dimensionless context, our derivation presented here offers three distinct contributions. First, we adopt a fundamentally different mathematical approach; rather than relying on complex Bessel function identities, we employ a direct integration method over spherical coordinates. Second, our formulation is explicitly dimensional, retaining physical parameters (e.g., $D$, $T$, $\overline{d}_0$, and $r_{\mathrm{Rx}}$) to seamlessly integrate with the system and signal models established in our earlier sections. Finally, our step-by-step derivation is physics-informed; by utilizing the geometric interaction between the molecular cloud and the sensing volume, it provides clear physical interpretations at each stage of the integration.

In the absence of drift, the concentration profile resulting from a point source release at $\mathbf{x}_0$ is given by
\begin{equation}
    c(\mathbf{x}, T) = \frac{1}{(4\pi DT)^{3/2}} \exp\left( -\frac{\|\mathbf{x}-\mathbf{x}_0\|^2}{4DT} \right).
\end{equation}
The sensing probability is defined as the integral of $c(\mathbf{x}, T)$ over the sensing region $\mathcal{R}_{\mathrm{s}} = \{\mathbf{x} \mid \|\mathbf{x}\| \leq r_{\mathrm{Rx}}\}$. By establishing a spherical coordinate system $(r, \theta, \phi)$ centered at the origin—where $r$ denotes the radial distance, $\theta \in [0, 2\pi)$ is the azimuthal angle, and $\phi \in [0, \pi]$ is the polar angle measured from the $x_1$-axis—and aligning the Tx at distance $\overline{d}_0$ along the $x_1$-axis, the distance term is expressed as $\|\mathbf{x}-\mathbf{x}_0\|^2 = r^2 + \overline{d}_0^2 - 2r\overline{d}_0\cos\phi$. The volume integral is formulated as
\begin{IEEEeqnarray}{rCl}
    p_s(T) &=& \frac{1}{(4\pi DT)^{3/2}} \int_0^{r_{\mathrm{Rx}}} \int_0^\pi \int_0^{2\pi} r^2 \sin\phi \nonumber \\
    && \times \exp\left( -\frac{r^2+\overline{d}_0^2-2r\overline{d}_0\cos\phi}{4DT} \right) d\theta d\phi dr. \quad \label{eq:app_b_vol_int}
\end{IEEEeqnarray}

We first evaluate the angular integrals. The integral over the azimuthal angle $\theta$ yields $2\pi$. For the polar angle $\phi$, letting $u = \cos\phi$ gives $\int_{-1}^1 \exp(\frac{r\overline{d}_0}{2DT} u) du = \frac{2DT}{r\overline{d}_0} \bigl( e^{\frac{r\overline{d}_0}{2DT}} - e^{-\frac{r\overline{d}_0}{2DT}} \bigr)$. Substituting this back into \eqref{eq:app_b_vol_int} yields a radial integral:
\begin{equation}
    p_{\mathrm{s}}(T) = \frac{1}{\sqrt{4\pi DT} \overline{d}_0} \int_0^{r_{\mathrm{Rx}}} r \left( e^{-\frac{(r-\overline{d}_0)^2}{4DT}} - e^{-\frac{(r+\overline{d}_0)^2}{4DT}} \right) dr.
\end{equation}
To solve the Gaussian-like integrals, we apply integration by parts. For the first term, we let $v = \frac{r - \overline{d}_0}{\sqrt{4DT}}$, yielding
\begin{equation}
\begin{aligned}
    &\int_0^{r_{\mathrm{Rx}}} r e^{-\frac{(r-\overline{d}_0)^2}{4DT}} dr = -2DT \left( e^{-\frac{(r_{\mathrm{Rx}}-\overline{d}_0)^2}{4DT}} - e^{-\frac{\overline{d}_0^2}{4DT}} \right) \\
    &\quad + \overline{d}_0\sqrt{\pi DT} \left[ \mathrm{erf}\left(\frac{r_{\mathrm{Rx}}-\overline{d}_0}{\sqrt{4DT}}\right) + \mathrm{erf}\left(\frac{\overline{d}_0}{\sqrt{4DT}}\right) \right].
\end{aligned}
\end{equation}
A similar procedure is applied to the second term involving $(r+\overline{d}_0)$. Upon combining the terms, the error function components containing only $\overline{d}_0$ cancel each other out. The final expression for the sensing probability is derived as
\begin{IEEEeqnarray}{rCl}
    p_{\mathrm{s}}(T) &=& \frac{1}{2} \left[ \mathrm{erf}\left( \frac{r_{\mathrm{Rx}}-\overline{d}_0}{\sqrt{4DT}} \right) + \mathrm{erf}\left( \frac{r_{\mathrm{Rx}}+\overline{d}_0}{\sqrt{4DT}} \right) \right] \nonumber \\
    && - \frac{\sqrt{DT}}{\overline{d}_0\sqrt{\pi}} \Biggl[ \exp\left( -\frac{(r_{\mathrm{Rx}}-\overline{d}_0)^2}{4DT} \right) \nonumber \\
    && \qquad \qquad - \exp\left( -\frac{(r_{\mathrm{Rx}}+\overline{d}_0)^2}{4DT} \right) \Biggr],
\end{IEEEeqnarray}
which matches the result presented in Section \ref{sec:analytical_CIR} and ensures physical consistency regarding dimensions and the non-monotonic nature of passive sensing.

\section{Field Design for Passive Receivers via Sensing Probability Maximization}
\label{appendix:MSP_design}

This appendix details the \textit{Maximize Sensing Probability} waveform design for the passive spherical receiver, demonstrating the universality of the MRP engine presented in Algorithm~\ref{alg:maxRP}. 

For the passive receiver, the received probability in the first slot is determined by the sensing probability at the sampling time $t_s$, i.e., $p_{\mathrm{R}}[1] = p_{\mathrm{s}}(t_s)$. From \eqref{eq:spherical_rx}, it can be shown that $p_{\mathrm{s}}(t_s)$ is a concave, monotonically decreasing function of the effective distance $\overline{d}_0(t_s)$. The maximum sensing probability is achieved when the effective distance is nullified, $\overline{d}_0(t_s) = 0$, which corresponds to physically shifting the center of the particle cloud to the origin of the Rx sphere.

To maximize $p_{\mathrm{s}}(t_s)$ using the Phase I signal-enhancing field $E_1^{\mathrm{S}}(t) = V_1$ applied during $t \in [0, t_s]$, we formulate the distance-minimization problem as follows, 
\begin{IEEEeqnarray}{rl}
    (\mathrm{P}_{\mathrm{PA}}) \quad \argmin_{V_1} & \quad \bigl\| x_0 + c_e V_1 t_s + U_1(t_s) \bigr\|^2 \nonumber \\
    \mathrm{s.t.} & \quad V_1^2 t_s \le \xi, \label{eq:app_c_opt1}
\end{IEEEeqnarray}
where $U_1(t_s) = \int_0^{t_s} u_1(\alpha) d\alpha$ is the background flow displacement. Solving for the root yields the ideal unconstrained velocity for Phase I as $q_{\mathrm{S}}^{\mathrm{P}} = -(x_0 + U_1(t_s))/(c_e t_s)$.

If residual energy $\xi_{\mathrm{res}} = \xi - (V_1^*)^2 t_s > 0$ remains, it is utilized in Phase II to apply an opposing field $V_2$ during $t \in [t_s, T_b]$ to suppress the leading ISI at the next sampling instance $t_s + T_b$. Since the field is periodic, the total displacement at $t_s + T_b$ includes the effects of $V_1^*$ applied twice (in both slots) and $V_2$ applied once. To aggressively deviate the particles away from the RX, we determine the optimal $V_2$ that maximizes the distance. Namely,
\begin{IEEEeqnarray}{rCl}
    \argmax_{V_2} && \quad \bigl\| x_0 + 2c_e V_1^* t_s + U_1(t_s + T_b) + c_e V_2 (T_b - t_s) \bigr\|^2 \nonumber \\
    \mathrm{s.t.} && \quad V_2^2 (T_b - t_s) \le \xi_{\mathrm{res}}. \label{eq:app_c_opt2}
\end{IEEEeqnarray}
Specifically, $q_{\mathrm{R}}^{\mathrm{P}} = -\bigl(x_0 + 2c_e V_1^* t_s + U_1(t_s + T_b)\bigr) / \bigl( c_e (T_b - t_s) \bigr)$ represents the target unconstrained velocity to nullify the cumulative displacement; its sign is utilized by the MRP engine to apply the maximum available suppression field in the opposite direction.

To ensure structural symmetry with the fully-absorbing design case, by directly inputting the parameters $q_{\mathrm{S}} = q_{\mathrm{S}}^{\mathrm{P}}$, $q_{\mathrm{R}} = q_{\mathrm{R}}^{\mathrm{P}}$, and explicitly mapping the temporal phases as $T_{p1} = t_s$, and $T_{p2} = T_b - t_s$ into the universal MRP Engine (Algorithm~\ref{alg:maxRP}), the optimal field magnitudes $(V_1^*, V_2^*)$ for the passive receiver are efficiently generated.

\balance
\bibliographystyle{IEEEtran}
\bibliography{tvaef}

\end{document}